\documentclass[%
 reprint,
superscriptaddress,
 amsmath,amssymb,
 aps,
]{revtex4-2}

\usepackage{graphicx}
\usepackage{dcolumn}
\usepackage{bm}
\usepackage{hyperref}

\DeclareMathAlphabet{\mathpzc}{OT1}{pzc}{m}{it}
 
\newcommand\beq{\begin{equation}}
\newcommand\eeq{\end{equation}}
\newcommand\bea{\begin{eqnarray}}
\newcommand\eea{\end{eqnarray}}
\newcommand\bwt{\begin{widetext}}
\newcommand\ewt{\end{widetext}}
\newcommand\nin{\noindent} 
\newcommand\nn{\nonumber\\}

\begin{document}


\title{Relativistic effects in k-essence}


\author{Didam G.A. Duniya}\email{duniyaa@biust.ac.bw}
\affiliation{Department of Physics \& Astronomy, Botswana International University of Science and Technology, Palapye, Botswana}

\author{Isaac Opio}\email{iopio2567@gmail.com}
\affiliation{Department of Physics \& Astronomy, Botswana International University of Science and Technology, Palapye, Botswana}

\author{Bishop Mongwane}\email{bishop.mongwane@uct.ac.za}
\affiliation{Department of Mathematics \& Applied Mathematics, University of Cape Town, South Africa}

\author{Hassan Abdalla}\email{hassanahh@gmail.com}
\affiliation{Centre for Space Research, North-West University, Potchefstroom 2520, South Africa}
\affiliation{Department of Astronomy and Meteorology, Omdurman Islamic University, Omdurman 382, Sudan}

\date{\today}

\begin{abstract}\nin
Relativistic effects are sensitive to subtle changes in dark energy. These effects grow on very large scales and at high redshifts, which will be the reach of upcoming surveys. We investigate these effects in both the linear and the angular galaxy power spectra in a late-time universe dominated by cold dark matter and k-essence, focusing on three core models (dilaton, tachyon, and DBI scalar fields) and contrasting their predictions with those of the concordance model. By enforcing identical present-day cosmological parameters, we isolate the imprints of k-essence dynamics and perturbations on very large scales. We found that relativistic corrections dominate on very large scales and grow with redshift, but are largely insensitive to k-essence microphysics in Fourier space, leading to strong degeneracies among the models. However, in the angular power spectrum, where line-of-sight integrals are naturally included, relativistic effects are significantly amplified, yielding better sensitivity to clustering k-essence. In particular, the tachyon exhibits clear deviations across multipoles and redshifts, with distinct imprints in the Doppler and the combined (velocity and gravitational) potentials contributions. Furthermore, our results show that neglecting relativistic corrections can lead to systematic misestimation of deviations of k-essence from the cosmological constant. Our results show the relativistic angular galaxy power spectrum as a more consistent and robust probe of ultra-large-scale physics. These findings underscore the need for full relativistic modelling in next-generation surveys that are targeting horizon-scale modes, where the imprint of non-standard dark energy is most pronounced.

\end{abstract}


\maketitle


\section{Introduction}
\label{sec:intro}

With precision cosmology era on the horizon, theoretical work needs to be done to prepare for implementing tests of models with data from the surveys. The upcoming surveys in the optical, infrared, and 21 cm emission of neutral hydrogen will probe vast cosmic volumes encompassing scales near or beyond the Hubble horizon. At these unprecedented scales and high redshifts, incorporating relativistic corrections (e.g.~\cite{Yoo:2009au,Yoo:2010ni, Yoo:2014kpa, Challinor:2011bk, Bonvin:2011bg, Jeong:2011as, Lopez-Honorez:2011emg, Alonso:2015uua, Bonvin:2014owa, Durrer:2016jzq, Duniya:2013eta, Duniya:2015nva, Duniya:2016ibg, Duniya:2015dpa, Duniya:2019mpr, Durrer:2016jzq}) into the large-scale power spectrum becomes essential. The relativistic corrections arise from several cosmic phenomena including Doppler, integrated Sachs-Wolfe (ISW), time-delay, and (gravitational and velocity) potentials effects. 

It is necessary to use a fully relativistic analysis if we are to extract maximal and accurate information from the next-generation large-volume surveys: the simplistic Newtonian calculation of the observed overdensity is inadequate on the given scales. These effects are known to be sensitive to small changes in the attributes of the underlying dark energy (e.g.~\cite{Duniya:2013eta, Armendariz-Picon:2000nqq, Armendariz-Picon:2000ulo, Piazza:2004df, Amendola:2010bk, Bamba:2012cp, Duniya:2015nva, Duniya:2015dpa, Tsujikawa:2010sc}) (or modified gravity, see e.g.~\cite{Clifton:2011jh, Ishak:2018his} for reviews). Moreover, for example, it is known that they act to boost the imprint of the given dark energy in large-scale cosmological observables (e.g.~\cite{Duniya:2016gcf, Duniya:2022vdi, Duniya:2022xcz, Duniya:2023xgx}); consequently, enhancing the potential of the given observables as cosmological probes. As pointed out in \cite{Duniya:2013eta}, relativistic effects become substantial on the same scales that are important for (1) testing dark energy models and modified gravity models, including general relativity itself, and (2) measuring the primordial non-Gaussianity signal in large scale. The sensitivity of relativistic effects to the underlying cosmological model will help in placing constraints on dark energy in the cosmological parameters, and will be important in distinguishing dark energy models. 

To our knowledge, there are no previous works in the literature that have probed relativistic effects in the linear and the angular galaxy power spectra for a universe with k-essence dark energy $\phi$ \cite{Armendariz-Picon:1999hyi, Chiba:1999ka, Armendariz-Picon:2000nqq, Armendariz-Picon:2000ulo, Brando:2020ouk, Amendola:2010bk, Bamba:2011ih, Tsujikawa:2012hv, Bamba:2012cp, Mohammadi:2019qeu, Rezazadeh:2020zrd, Mulki:2020hwt} (``k-essence," henceforth). It may be worth noting that k-essence, with Lagrangian $P(\phi, X)$ and kinetic term $X(\dot{\phi})$ where $\dot{\phi}$ is time derivative of $\phi$, was originally introduced as ``k-inflation'' in \cite{Armendariz-Picon:1999hyi}, as a dynamical explanation for the well known early-time cosmic inflation. The first application of k-essence to the late-time cosmic acceleration, as dark energy, was done in \cite{Chiba:1999ka}, where it was referred to as ``kinetically driven quintessence.'' It was further presented, for the first time as ``k-field'' or ``k-essence'' in \cite{Armendariz-Picon:2000nqq}, as a dynamical solution to the problem of a small cosmological constant and late-time cosmic acceleration. Later, its present name was settled for in \cite{Armendariz-Picon:2000ulo}, where its essentials (basics and dynamical analysis) were presented.

For works on matter perturbations, matter linear power spectrum, and matter growth rate in k-essence see e.g.~\cite{Brando:2020ouk, Bamba:2011ih, Tsujikawa:2012hv, Mohammadi:2019qeu, Rezazadeh:2020zrd, Mulki:2020hwt}. It should be pointed out that in \cite{Brando:2020ouk}, ``relativistic effects'' is used in a different context, where it is understood to refer to effects owing to the presence of massless neutrinos and photons in the evolution equations, and not owing to Doppler effect, ISW effect, time-delay, and potentials effect, in the observed galaxy overdensity. In this paper, we investigate relativistic effects in both the linear and the angular galaxy power spectra, respectively. We perform our analysis for a late-time universe dominated by k-essence and cold dark matter (kCDM), using three well studied k-essence models. 

The purpose of this study is not to perform a parameter fit for the kCDM models, but to show how relativistic effects in these models differ from those in the cosmological concordance model ($\Lambda$CDM) on very large scales; demonstrating the physical divergence of k-essence from the cosmological constant on the largest scales. Thus, we choose parameters so that the kCDM models have the same present-day value of the matter density parameter $\Omega_{m0}$ and Hubble constant $H_0$ as $\Lambda$CDM. This ``normalizes'' the power spectra to be the same on small scales, at the present epoch. The advantage of this is that all deviations owing to k-essence clustering and relativistic corrections become isolated on very large scales.

In what follows, we discuss the kCDM cosmological equations, and analyse the evolutions of the equation of state parameters and sound speeds, in \S\ref{sec:kCDM}. We outline the observed number-count overdensity of galaxy redshift surveys in \S\ref{sec:Delta_n}. We compute and analyse the linear galaxy power spectrum in \S\ref{sec:Pks}. We ensure that the stability conditions (e.g.~\cite{Amendola:2010bk}) $\partial^2 P(\phi, X) / \partial{X}^2 \,{>}\, 0$ and $0 \,{\leq}\, c_{s\phi}^2 \,{\leq}\, 1$ are satisfied, where $c_{s\phi}$ is the effective speed of $\phi$ perturbations. In \S\ref{sec:Cls} we compute and analyse the angular galaxy power spectrum. We give a detailed conclusion in \S\ref{sec:Concl}.

\section{The kCDM Cosmological Equations}
\label{sec:kCDM}
We consider a late-time kCDM model, with k-essence being given by a scalar field $\phi$ whose dynamics is driven solely by (non-canonical) kinetic terms, i.e. high-order derivatives of $\phi$. (See cited references in \S\ref{sec:intro} for further details on k-essence.)

In general, the action for kCDM models is given by
\beq\label{action}
S = \int d^4x \sqrt{-g} \left[\dfrac{1}{2\kappa^2}R + P(\phi,X)\right] + S_m,
\eeq
where $\kappa\, {\equiv}\, \sqrt{8\pi G}$, with $G$ being the Newton's gravitational constant (we take the speed of light $c \,{=}\, 1$); $g$ is the determinant of the metric tensor, $R$ is the well known Ricci scalar, and $P(\phi,X)$ is the Lagrangian (density) for k-essence; with $X$ measuring the kinetic energy, given by
\beq\label{X-defn}
X \equiv -\dfrac{1}{2} \partial^\mu \phi \partial_\mu \phi,
\eeq
where $\partial_\mu \,{=}\, \partial/\partial{x}^\mu$ (assuming standard index raising and lowering conventions), and the indices $\mu \,{=}\, 0,\, 1,\, 2,\, 3$. The quantity, $S_m$, is the action for (cold) dark matter. In a kCDM, the kinetic energy \eqref{X-defn} is the source of the cosmic acceleration, in the late times.

From the kCDM action \eqref{action}, we obtain the field equations, given by 
\beq\label{Einstein}
R_{\mu\nu} - \dfrac{1}{2} g_{\mu\nu} R = \kappa^2 \left(T^m_{\mu\nu} + T^\phi_{\mu\nu}\right),
\eeq
where $g_{\mu\nu}$ is the metric tensor, $R_{\mu\nu}$ is the Ricci tensor, $T^m_{\mu\nu}$ is the matter energy-momentum tensor resulting from the variation of $S_m$ with respect to $g^{\mu\nu}$, and $T^\phi_{\mu\nu}$ is the energy-momentum tensor for k-essence, given by
\beq\label{Tmunu-kess}
T^\phi_{\mu\nu} = -\dfrac{2}{\sqrt{-g}} \dfrac{\delta(\sqrt{-g} P)}{\delta g^{\mu\nu}} = \partial_X P\, \partial_\mu \phi \partial_\nu\phi + g_{\mu\nu} P,
\eeq
where $\partial_X \,{=}\, \partial/\partial{X}$, and $P$ is as given in \eqref{action}. 

The k-essence energy-momentum tensor \eqref{Tmunu-kess} shows that the k-essence Lagrangian essentially corresponds to an effective ``pressure''---with both $T^m_{\mu\nu}$ and $T^\phi_{\mu\nu}$ describing the energy and momentum of perfect fluids, in the form given by
\beq\label{Tmunu-A}
T^A_{\mu\nu} = \left(\rho_A + P_A\right) u^A_\mu u^A_\nu + g_{\mu\nu}P_A,
\eeq
where $\rho_A$ are the energy densities for dark matter ($A \,{=}\, m$) and k-essence ($A \,{=}\, \phi$), $P_A$ are the pressures, and $u^A_\mu$ are the 4-velocities, accordingly; taking the k-essence energy density and 4-velocity as
\begin{align}\label{rho_kess}
\rho_\phi \equiv &\; 2X\partial_X P_\phi - P_\phi,\\
\label{u_kess}
u^\phi_\mu \equiv &\; \dfrac{\partial_\mu \phi}{\sqrt{2X}}, \quad P_\phi(X) = P(\phi,X),
\end{align}
with $P(\phi,X)$ as in \eqref{action}.


\subsection{General background equations}
\label{subsec:background}
By using \eqref{Einstein} and \eqref{Tmunu-A}, the Friedmann equation of the background Universe, is given by 
\beq\label{Friedmann}
{\cal H}^2 = \dfrac{\kappa^2 a^2}{3}\left(\rho_m + \rho_\phi\right),
\eeq
where $\rho_\phi$ is as given by \eqref{rho_kess}, ${\cal H} \,{=}\, a'/a$ is the comoving Hubble parameter, $a \,{=}\, a(\eta)$ is the cosmic scale factor and, a prime denoting derivative with respect to conformal time $\eta$.

The evolution of the Hubble parameter is governed by the equation
\begin{align}\label{acceleration}
{\cal H}' =&\; -\dfrac{\kappa^2 a^2}{6} \left(\rho_m + 2P_\phi + 2X\partial_X P_\phi\right), \nn
=&\; -\dfrac{1}{2} {\cal H}^2 \left(1 + 3w\right),
\end{align}
where $w$ is the effective equation of state parameter of the system; with $P$ being the total pressure (dark matter being pressureless, $P_m \,{=}\, 0$), and $\rho$ the total energy density. The equation of state parameters associated with the energy-momentum tensors \eqref{Tmunu-A}, are given by
\beq\label{w_A}
w_A \equiv \dfrac{P_A}{\rho_A},
\eeq
where $w_m \,{=}\, 0$ (henceforth), and 
\beq\label{w_phi}
w_\phi = \dfrac{P_\phi}{2X\partial_X P_\phi - P_\phi} ,
\eeq
provided $P_\phi \,{\neq}\, 2X\partial_X P_\phi$, where we used \eqref{rho_kess}. This gives the general form of the equation of state parameter for general k-essence. We see from \eqref{w_phi} that, when $|2X\partial_X P_\phi| \ll |P_\phi|$ we have $w_\phi \,{\to}\, {-}1$ (value in $\Lambda$CDM).

The conservation of the total energy-momentum tensor, implies
\beq\label{conservation}
\nabla_\mu T_A^{\mu\nu} = 0,
\eeq
where we assume (henceforth) that dark matter and k-essence interact only gravitationally. Thus, in the background, the total energy-momentum tensor conservation \eqref{conservation} leads to 
\begin{align}\label{rhoDots}
\rho'_A + 3{\cal H}(1 + w_A) \rho_A = 0,
\end{align}
where $w_A$ is as given by \eqref{w_A}. By using \eqref{rho_kess} in \eqref{rhoDots}, we have the general Klein-Gordon equation for k-essence, given by 
\beq\label{KleinGordon}
\phi'' + 2 {\cal H} \Gamma (\partial_X P_\phi - X\partial^2_X P_\phi) \phi' = a^2 \Gamma (\partial_\phi P_\phi - 2X \partial^2_{X\phi} P_\phi),
\eeq
where,
\beq\label{Gamma}
\Gamma^{-1} \equiv \partial_X P_\phi + 2X \partial^2_X P_\phi,
\eeq
with $\partial_\phi \,{=}\, \partial/\partial\phi$, $\partial^2_{X\phi} \,{=}\, \partial^2/(\partial X\partial\phi)$ and, $\partial_X$ being as in \eqref{Tmunu-kess}. The physical sound speed, $c_{s\phi}$, is given by
\beq\label{cs2}
c^2_{s\phi} = \dfrac{\partial_X P_\phi}{\partial_X \rho_\phi}.
\eeq
In what follows (\S\ref{subsec:dilaton}--\S\ref{subsec:DBI}), we give the particular equations for three k-essence models.


\subsection{Dilatonic ghost condensate}\label{subsec:dilaton}
The dilatonic ghost condensate dark energy \cite{Piazza:2004df, Amendola:2010bk} (``dilaton,'' henceforth) is a string-theory inspired model. The effective pressure (Lagrangian) of the dilaton, is given by
\beq\label{dilaton-P}
P_\phi = -X + c_1X^2 e^{\lambda\kappa\phi},
\eeq
where $c_1 \,{=}\, 1/M^4$ with $M$ (a constant) being a mass scale, $\kappa$ and $X$ are as given in \eqref{action} and \eqref{X-defn}, respectively, and $\lambda$ is a dimensionless parameter. By using \eqref{dilaton-P} in \eqref{rho_kess}, we have the energy density for the dilaton, given by
\beq\label{dilaton-rho}
\rho_\phi = -X + 3c_1X^2 e^{\lambda\kappa\phi}.
\eeq
By using \eqref{dilaton-P} and \eqref{dilaton-rho} in \eqref{w_phi}, we have the equation of state parameter of dilaton, given by
\beq\label{dilaton-w}
w_\phi = \dfrac{1 - c_1 a^{-2} \phi'^2 e^{\lambda\kappa\phi}/2}{1 - 3c_1 a^{-2} \phi'^2 e^{\lambda\kappa\phi}/2}, 
\eeq
where we have from \eqref{X-defn}. The particular Klein-Gordon equation for the dilaton, is given by
\beq\label{dilaton-KG}
\phi'' \left(1 - 3c_1\dfrac{\phi'^2}{a^2} e^{\lambda\kappa\phi}\right) + 2{\cal H}\phi' - 3c_1\lambda\dfrac{\kappa\phi'^4}{4a^4} e^{\lambda\kappa\phi} = 0
\eeq
where we used the effective pressure \eqref{dilaton-P} in \eqref{KleinGordon}. (Notice the error in \cite{Amendola:2010bk}, equation (8.37).) The sound speed for the dilaton, is given by
\beq\label{dilaton-cs2}
c^2_{s\phi} = \dfrac{1 - c_1 a^{-2} \phi'^2 e^{\lambda\kappa\phi} }{1 - 3c_1 a^{-2} \phi'^2 e^{\lambda\kappa\phi}}, 
\eeq
which dictates how pressure perturbations propagate in a universe with dilaton as dark energy.


\subsection{Tachyon field}\label{subsec:tachyon}
In this work, we consider the tachyon field (``tachyon,'' henceforth) as k-essence because it falls into a class of the action \eqref{action}. The tachyon arises from low-energy effective theory of D-branes and open strings. 

The effective pressure for tachyon is given by
\beq\label{tachyon-P}
P_\phi = -U\sqrt{1 - 2c_1X},
\eeq
where $U \,{=}\, U(\phi)$ is a potential of tachyon. Notice the factor of $1/M^4$ in \eqref{tachyon-P}. In \S\ref{subsec:dilaton}, by the definition of the effective pressure \eqref{dilaton-P}, it implies that $\phi$ acquires the dimension of mass when $\kappa \,{\neq}\, 1$ and $c \,{=}\, 1$ (as assumed in this work). However, for tachyon if the given factor is omitted (as commonly done in the literature) it will imply that $\phi$ will have a dimension of per mass. Thus, for consistency in the dimension of $\phi$, we include the given factor in the tachyon effective pressure \eqref{tachyon-P}, to give $\phi$ the dimension of mass as in \S\ref{subsec:dilaton}. 

Similarly, given \eqref{rho_kess} and \eqref{tachyon-P}, we have the energy density of dilaton, given by
\beq\label{tachyon-rho}
\rho_\phi = \dfrac{U}{\sqrt{1-2c_1X}},
\eeq
and, given \eqref{w_phi}, \eqref{tachyon-P} and \eqref{tachyon-rho}, we have
\beq\label{tachyon-w}
w_\phi = -1 + c_1\dfrac{\phi'^2}{a^2},
\eeq
which gives the equation of state parameter of tachyon. The associated Klein-Gordon equation for tachyon, is given by
\beq\label{tachyon-KG}
\phi'' + 2{\cal H}\phi' \left(1 - \dfrac{3c_1\phi'^2}{2a^2}\right) + \dfrac{a^2}{c_1} \left(1 - c_1\dfrac{\phi'^2}{a^2} \right) \partial_\phi \ln{U} = 0,
\eeq
where we have used \eqref{KleinGordon} and \eqref{tachyon-P}. Note that \eqref{tachyon-w} can be used to eliminate the $\phi'^2$ terms in the brackets in \eqref{tachyon-KG}.

The physical sound speed, is given by
\beq\label{tachyon-cs2}
c^2_{s\phi} = 1 - c_1\dfrac{\phi'^2}{a^2},
\eeq
which gives the squared speed of pressure perturbations in a universe dominated by tachyon and dark matter.

In order for tachyon to account for the late-time cosmic acceleration, its potential needs to be shallower than the inverse squared potential \cite{Amendola:2010bk}, i.e. the potential takes the form $U(\phi) \,{\propto}\, \phi^{-\alpha}$ where $0 \,{<}\,\alpha \,{<}\, 2$. Here we take
\beq\label{tachyon-U}
U(\phi) = \dfrac{M^{4+\alpha}}{\phi^\alpha}, 
\eeq
with $\alpha$ being a constant, and $M$ a (constant) mass scale. Note that both $w_\phi$ and $c_{s\phi}$ are independent of $U(\phi)$; however, in order to compute them we need to solve the $\phi$ equation of motion \eqref{tachyon-KG}. Hence $w_\phi$ and $c_{s\phi}$ are indirectly dependent on $U(\phi)$.


\subsection{Dirac-Born-Infeld field}\label{subsec:DBI}
The Dirac-Born-Infeld (DBI) field is another model that belongs to the class of the Lagrangian \eqref{action}. The Lagrangian density for the DBI field, is given in general form by (e.g. \cite{Amendola:2010bk})
\beq\label{DBI-P1}
P(\phi,X) = -f(\phi)^{-1}\sqrt{1-2f(\phi)X} + f(\phi)^{-1} - U(\phi),
\eeq
where $U(\phi)$ is the DBI field potential, and $f(\phi)$ is a coupling factor. In this work we will only consider a special case of \eqref{DBI-P1}, given by \cite{Armendariz-Picon:2000nqq}
\beq\label{DBI-P2}
P(\phi,X) = \dfrac{M^6}{\phi^2} \tilde{P}(X) \equiv P_\phi,
\eeq
where,
\beq\label{DBI-P3}
\tilde{P}(X) = -2.01 + 2\sqrt{1 + c_1 X} + 0.03(c_2 X)^3 - (c_3 X)^4 ,
\eeq
with $c_1$ as in \eqref{dilaton-P}, $c_2 \,{=}\, 10^{-5} c_1$ and $c_3 \,{=}\, 10^{-6} c_1$. Consequently, $\tilde{P}$ is dimensionless. (See also~\cite{Amendola:2010bk, Armendariz-Picon:2000ulo}, for a different expression of $\tilde{P}$.) The (constant) factors $M^6$ and $M^{-4}$ in \eqref{DBI-P2} and \eqref{DBI-P3}, respectively, are introduced for consistency in the dimension of $\phi$: they ensure that the DBI field has the dimension of mass as the dilaton and the tachyon (\S\ref{subsec:dilaton} and \S\ref{subsec:tachyon}). Note that in the literature there are works, e.g. \cite{Armendariz-Picon:2000nqq}, where $M \to 1$ and $P_\phi \,{=}\, \tilde{P}/\phi^2$, with $c_1 \,{=}\, 1$, $c_2 \,{=}\, 10^{-5}$ and $c_3 \,{=}\, 10^{-6}$. However, this is only because the authors assumed that $\kappa \,{=}\, 1$, which is not particularly necessary.

The transformation \eqref{DBI-P2} implies that,
\begin{align}\label{DBI-rho1}
\rho_\phi =&\; \left(2X\partial_X \tilde{P} - \tilde{P}\right) M^6 / \phi^2,\nn
=&\; \dfrac{M^6}{\phi^2} \left[2.01 - \dfrac{2}{\sqrt{1 + c_1 X}} + 0.15(c_2 X)^3 - 7(c_3 X)^4\right].
\end{align}
The equation of state parameter for the DBI field, is 
\beq\label{DBI-w}
w_\phi = \dfrac{-2.01 + 2\sqrt{1+c_1 X} + 0.03(c_2 X)^3 - (c_3 X)^4}{2.01 - 2 (1+c_1 X)^{-1/2} + 0.15(c_2 X)^3 - 7(c_3 X)^4},
\eeq
and, given \eqref{KleinGordon}, \eqref{DBI-P2} and \eqref{DBI-P3}, the particular Klein-Gordon equation for the DBI field, is given by
\begin{align}
\phi'' +&\; 2{\cal H}\phi'\Gamma_x \dfrac{M^6}{\phi^2} \Big[\dfrac{1}{2}c_1(2+3c_1X) (1+c_1X)^{-3/2} \nn
&\hspace{1.75cm} - 0.09c_2(c_2X)^2 + 8c_3(c_3X)^3\Big] \nn
=&\; 2a^2\Gamma_x \dfrac{M^6}{\phi^3} \Big[2.01 - \dfrac{2}{\sqrt{1+c_1X}} + 0.15(c_2X)^3 \nn
&\qquad\qquad - 7(c_3X)^4 \Big],
\end{align}
where, given \eqref{Gamma}, we have
\beq
\dfrac{\phi^2}{M^6} \Gamma_x^{-1} = \dfrac{c_1}{(1+c_1 X)^{3/2}} + 0.45c_2(c_2X)^2 - 28c_3(c_3X)^3.
\eeq
The sound speed for the DBI field, is given by
\beq\label{DBI-cs2}
c^2_{s\phi} = \dfrac{c_1(1+c_1 X)^{-1/2} + 0.09c_2(c_2 X)^2 - 4c_3(c_3 X)^3}{c_1(1+c_1 X)^{-3/2} + 0.45c_2(c_2 X)^2 - 28c_3(c_3 X)^3}.
\eeq
This measures the propagation speed of pressure perturbations in kCDM, with the DBI field as dark energy.


\subsection{Perturbations equations}
\label{subsec:Perturbations}

\begin{figure*}\centering
\includegraphics[scale=0.4]{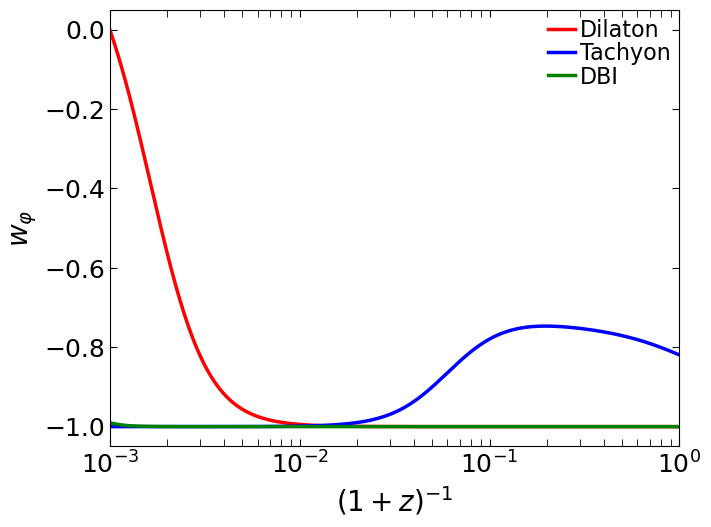} \includegraphics[scale=0.4]{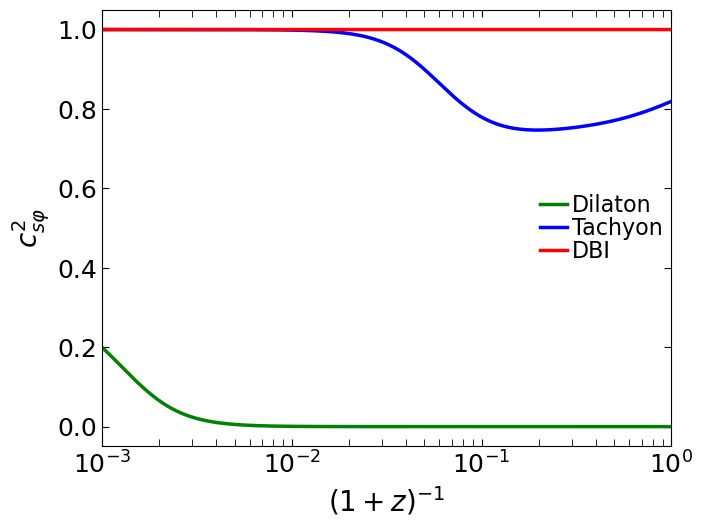}
\caption{\emph{Left}: The plots of the cosmic evolution of k-essence equation of state parameters \eqref{dilaton-w}, \eqref{tachyon-w}, and \eqref{DBI-w}, for dilaton, tachyon, and DBI, respectively---with $z$ being the cosmological redshift. \emph{Right}: The plots of the cosmic behaviour of k-essence physical sound speeds (squared) \eqref{dilaton-cs2}, \eqref{tachyon-cs2}, and \eqref{DBI-cs2}, for dilaton, tachyon, and DBI, respectively.}
\label{fig:wx+cs2}
\end{figure*}

Henceforth, we adopt the spacetime metric, given by
\beq\label{metric}%
ds^2 = a^2\left[-(1+2\Phi)d{\eta}^2 + (1-2\Phi)d\vec{x}\,^2\right],
\eeq
with $\Phi$ being the (gauge-invariant) Bardeen potential \cite{Bardeen:1980kt, Bonvin:2011bg, Duniya:2016ibg}, which is gravitationally prescribed by the relativistic Poisson equation, given by
\beq\label{Poisson} 
\nabla^2\Phi={3\over 2}{\cal H}^2(\Omega_m\Delta_m+\Omega_\phi\Delta_\phi),
\eeq
where $\Delta_A$ are the {\em comoving} density contrasts, defined by
\beq\label{D_A}
\Delta_A \equiv \delta_A - 3(1+w_A){\cal H} V_A, 
\eeq
where $\delta_A \,{\equiv}\, \delta\rho_A/\rho_A$ is the coordinate density contrast, $\delta\rho_A$ is the coordinate density perturbation, and $V$ is the peculiar velocity potential. It should be pointed out that \eqref{D_A} applies to each species, i.e. the comoving density contrasts are in the separate comoving gauge of each species (dark matter and k-essence), accordingly.

On very large scales (as considered in this work), it is important to use $\Delta_A$ instead of $\delta_A$, for the analysis. This is because on large scales $\delta_A$ is no longer an accurate tracer of the potential, and the relativistic Poisson equation \eqref{Poisson} must also be used. As pointed out in \cite{Duniya:2013eta}, although it is common in the literature to neglect $\Delta_\phi$ and use $\Delta_m \approx \delta_m$, which leads to the (limited) Newtonian Poisson equation, $\nabla^2\Phi \,{=}\, (3/2){\cal H}^2\Omega_m\delta_m$, such a setup is only valid on small (sub-Hubble) scales.

The potential evolution is driven by the total momentum density, given by
\beq\label{dPhidt}
\Phi' + {\cal H}\Phi = -{3\over 2}{\cal H}^2\left[\Omega_m V_m +(1+w_\phi)\Omega_\phi V_\phi\right].
\eeq
The dark matter fluctuations obey energy and momentum conservations, that lead to
\begin{align}\label{dDmdt}
\Delta'_m - {9\over 2}{\cal H}^{2}\Omega_\phi(1+w_\phi)\left(V_m-V_\phi\right) =& -\nabla^2V_m, \\
\label{dVmdt}
V'_m + {\cal H}V_m =& -\Phi,
\end{align}
and similarly for k-essence, we have
\begin{align}
\Delta'_\phi - 3w_\phi{\cal H}\Delta_\phi -& \frac{9}{2}{\cal H}^{2}\Omega_m(1+w_\phi)\left(V_\phi-V_m\right)  \nonumber\\ \label{dDxdt}
 =& -(1+w_\phi)\nabla^2V_\phi,\\ \label{dVxdt}
 V'_\phi + {\cal H}V_\phi =& -\Phi - \dfrac{c_{s\phi}^2}{(1+w_\phi)}\Delta_\phi ,
\end{align}
where $c_{s\phi}$ is the k-essence sound speed, given in \S\ref{subsec:background}-\S\ref{subsec:DBI}. Note that \eqref{Poisson}--\eqref{dVxdt} hold for any form of dark energy: it will only require one to specify $w_\phi$ and $c_{s\phi}$.


\subsection{Cosmic evolutions of $w_\phi$ and $c_{s\phi}$}
\label{subsec:wx-cs2}
We initialize evolutions at the epoch of (photon-matter) decoupling, $a_d \,{=}\, 1/(1 \,{+}\, z_d) \,{=}\, 0.001$, with $z_d$ being the redshift at decoupling. To determine the initial background densities of matter and k-essence, we set all the models to give the same present-day value of matter density parameter and Hubble constant: $\Omega_{m0} \,{=}\, 0.3$ and $H_0 \,{=}\, 67.8\, {\rm km\cdot s^{-1}\cdot Mpc^{-1}}$, respectively (consistent with $\Lambda$CDM Planck \cite{Planck:2018vyg} results). The advantage of this is that the power spectra coincide at today (redshift $z \,{=}\, 0$), and any deviations arising from the clustering of k-essence are isolated on large scales (the scales of interest in this work). For the dilaton, we set ${\kappa^2 M^4} \,{=}\, {3H_0^2}$ and $\lambda \,{=}\, {-}10^{-8}$, we take ${\kappa^{2+\alpha} M^{4+\alpha}} / {(3H_0^2)} \,{=}\, 44.856191$ and $\alpha \,{=}\, 0.5$ for the tachyon, and we use ${\kappa^2 M^4}/{(3H_0^2)} \,{=}\, 10^9$ for the DBI field.

In Fig.~\ref{fig:wx+cs2} (left panel) we show the cosmic evolutions of k-essence equation of state parameters \eqref{dilaton-w}, \eqref{tachyon-w}, and \eqref{DBI-w} for the dilaton, the tachyon, and the DBI field, respectively. We see that the dilaton is in the tracking regime at decoupling, with an equation of state parameter equal to that of the dominant component (matter). This suggests that in order to match current $\Lambda$CDM background cosmological parameter values from the cosmic microwave background (CMB) anisotropy measurements, the dilaton will need to assume a tracking behaviour at the decoupling epoch. Moreover, we see that it quickly evolves into a constant solution, behaving like a cosmological constant at $a \,{\gtrsim}\, 10^{-2}$ with $w_\phi \,{\simeq}\, {-}1$. This can be understood to be caused by early high Hubble rate or rapid cosmic expansion, which causes the drag or damping term, $2{\cal H}\phi' / (1-3c_1 a^{-2} \phi'^2 e^{\lambda \kappa \phi})$, in the dilaton equation of motion \eqref{dilaton-KG} to rapidly diminish the evolution rate $\phi'$ as the scale factor $a$ increases (expansion). Consequently, this effectively drags the amplitude of the dilaton to a fixed value, or $\phi' (a \,{>}\,a_d)\,{\to}\, 0$, till late times. Similarly for the DBI field, except that the DBI field shows relatively much lower tracking behaviour, and is already nearly fixed to a constant solution at decoupling by the rapid Hubble rate. In other words, the damping term in the DBI-field equation of motion \eqref{DBI-w} will remain effectively large throughout its evolution, suppressing the amplitude of the DBI field to a fixed value. Thus, although it is inherently dynamic, the DBI field will need to remain nearly static soon after decoupling until the present day in order to match current $\Lambda$CDM cosmological parameters from CMB; whereas, the dilaton will need to become static at $a \,{\gtrsim}\, 10^{-2}$. The sudden departure from a time-varying state to a time-independent or static state is widely known as ``freezing.'' Thus, essentially, the dilaton and the DBI field will need to be both tracking and freezing to satisfy chosen requirements for the background cosmology. By contrast, the tachyon behaves like a ``thawing'' dark energy: at the decoupling epoch it becomes effectively frozen in place, during which period it mimics the cosmological constant, until at $a\,{>}\,10^{-2}$ when it departs from this static state and evolves with the cosmic expansion. (See e.g.~\cite{Tsujikawa:2010sc}, for freezing and thawing quintessence.)

Similarly, in Fig.~\ref{fig:wx+cs2} (right panel) we show the plots of the cosmic evolution of k-essence squared sound speeds \eqref{dilaton-cs2}, \eqref{tachyon-cs2}, and \eqref{DBI-cs2} for the dilaton, the tachyon, and the DBI field, respectively. We see that for all the k-essence fields, the sound speeds (squared) are restricted as $0 \,{\leq}\, c^2_{s\phi} \,{\leq}\, 1$, with the lower bound only matched by the dilaton. By comparing \eqref{dilaton-w} and \eqref{dilaton-cs2}, we obtain
\beq\label{dil-cs2-w}
c^2_{s\phi} = (1+w_\phi) / (5-3w_\phi),\quad ({\rm Dilaton})
\eeq
so that when $w_\phi \,{\simeq}\, 0$ we have $c^2_{s\phi} \,{\simeq}\, 0.2$, and when $w_\phi \,{\simeq}\, {-}1$ we have $c^2_{s\phi} \,{\simeq}\, 0$. This is seen in the given figure. For the tachyon, the sound speed behaviour can easily be obtained by comparing \eqref{tachyon-w} and \eqref{tachyon-cs2}: 
\beq\label{tac-cs2-w}
c^2_{s\phi} = {-}w_\phi.\quad ({\rm Tachyon})
\eeq
Thus, we get for $c^2_{s\phi}$ the reflection of the behaviour of $w_\phi$, as shown by the results. However, for the DBI field, expressing $c^2_{s\phi}$ in terms of $w_\phi$ is not analytically tractable since both quantities are highly nonlinear (square roots, inverse powers, cubic and quartic terms). Nevertheless, by the evolution of the equation of state parameter of the DBI field (FIG.~\ref{fig:wx+cs2}, right panel), it is reasonable to assume that the kinetic term $X$ remains in slowly varying regimes, and hence having weak higher-order terms. By this, we can combine Taylor expansion and iterative perturbation method on $w_\phi$ and $c^2_{s\phi}$, to get
\beq\label{dbi-cs2-w}
c^2_{s\phi} \approx 1 + (w_\phi + 1)/200,\quad ({\rm DBI})
\eeq
and when $w_\phi \,{\simeq}\, {-}1$ we have $c^2_{s\phi} \,{\simeq}\, 1$, as seen in the plots. Thus, for the DBI field, sound travels almost at luminal speed. Note that corrections in \eqref{dbi-cs2-w} will grow with variations in $X$ or higher-order terms, e.g. if $X^3 \ll X^2 < 1$, then there will be a correction term $\propto (w_\phi + 1)^2$; thereby extending the linear approximation (see Appendix~\ref{App:DBI_wx_cs2_expand}). 

It should be emphasized that, although our normalization may affect the evolution of $w_\phi$, it does not determine the relationship between $c^2_{s\phi}$ and $w_\phi$: this is fixed by the Lagrangian. Equations \eqref{dil-cs2-w}, \eqref{tac-cs2-w}, and \eqref{dbi-cs2-w} are analytical and completely independent of the cosmological setup. However, for the dilaton and the tachyon, the given $c^2_{s\phi}(w_\phi)$ expressions, \eqref{dil-cs2-w} and \eqref{tac-cs2-w}, are exact while the expression \eqref{dbi-cs2-w} for the DBI field is approximate, which is owing to the highly nonlinear nature of $c^2_{s\phi}$ and $w_\phi$ of the DBI field.


\section{The Observed Galaxy overdensity}
\label{sec:Delta_n}
The galaxy number counts allows cosmologists to probe the growth of structure, and the nature of dark energy, on ultra-large scales (i.e. near and beyond the Hubble horizon). The observed distribution of galaxies in a given direction ${-}{\bf n}$ at redshift $z$, inherently acquires perturbations in the galaxy number-count density, in an inhomogeneous universe.

The observed, relativistic galaxy number overdensity (e.g.~\cite{Yoo:2009au, Yoo:2010ni, Yoo:2014kpa, Challinor:2011bk, Bonvin:2011bg, Jeong:2011as, Lopez-Honorez:2011emg, Alonso:2015uua, Bonvin:2014owa, Bonvin:2015kuc, Gaztanaga:2015jrs, Duniya:2016ibg, Durrer:2016jzq}) as measured by galaxy redshift surveys, is given by
\bea\label{Delta_n}
\Delta^{\rm obs}_{\rm g}({\bf n},z_S) \;=\; \Delta^{\rm std}_{\rm g}({\bf n},z_S) + \Delta^{\rm rels}_{\rm g}({\bf n},z_S),
\eea
where we take the standard term to be given by
\begin{align}\label{stdDelta_n}
\Delta^{\rm std}_{\rm g}({\bf n},z_S) \equiv &\; \Delta_{\rm g}({\bf n},z_S) + \dfrac{1}{{\cal H}}\, \partial_r \left({\bf n} \,{\cdot}\, {\bf V}\right) ({\bf n},z_S) \nn
&+ 2\int^{\bar{r}_S}_0{d\bar{r} \left(\bar{r} - \bar{r}_S\right)\dfrac{\bar{r}}{\bar{r}_S} \nabla^2_\perp \Phi(k,r) } ,\;
\end{align}
where $\Delta_{\rm g}$ is the comoving galaxy number overdensity, $\bar{r}_S \,{=}\, \bar{r}(z_S)$ is the background comoving distance at the source redshift ($z \,{=}\, z_S$), and $\nabla^2_\perp \,{=}\, \nabla^2 - \partial_r^2 - 2\bar{r}^{-1} \partial_r$ is the Laplacian on the plane transverse to the line of sight (the other terms retaining their standard notations). The first, the second and third terms, on the right hand side in \eqref{stdDelta_n}, give the density amplitude, the well-known redshift-space-distortions correction, and the (integral) weak-lensing correction, respectively.

The relativistic term in the observed number-count overdensity \eqref{Delta_n}, is given by
\begin{align}\label{relsDelta_n}
\Delta^{\rm rels}_{\rm g} ({\bf n},z_S) \equiv & \left(\dfrac{{\cal H}'}{{\cal H}^2} + \dfrac{2}{{\cal H}\bar{r}_S} -b_e \right) \left({\bf n} \,{\cdot}\, {\bf V}\right) ({\bf n},z_S) \nn
&+\; 2\left(\dfrac{{\cal H}'}{{\cal H}^2} + \dfrac{2}{{\cal H}\bar{r}_S} - b_e\right) \int^{\bar{r}_S}_0{d\bar{r} \Phi'(k,r) } \nn
&+\; \left(3 - b_e\right){\cal H}V({\bf n},z_S) + \dfrac{1}{{\cal H}}\Phi'({\bf n},z_S) \nn
&+ \left(\dfrac{{\cal H}'}{{\cal H}^2} + \dfrac{2}{{\cal H} \bar{r}_S} -b_e -1\right) \Phi({\bf n},z_S) \nn
&+\; \dfrac{4}{\bar{r}_S}\int^{\bar{r}_S}_0{d\bar{r} \Phi(k,r)}, 
\end{align}
where the first line gives the Doppler-effect correction, the second line gives the correction for ISW effect, the third and the fourth lines correct for (local) velocity-potential and gravitational-potential effects, accordingly, and the last line corrects for time-delay. These constitute the ``relativistic corrections'' in the redshift number-count overdensity.


\section{The Linear Power Spectrum}
\label{sec:Pks}

\begin{figure*}\centering
	\includegraphics[scale=0.33]{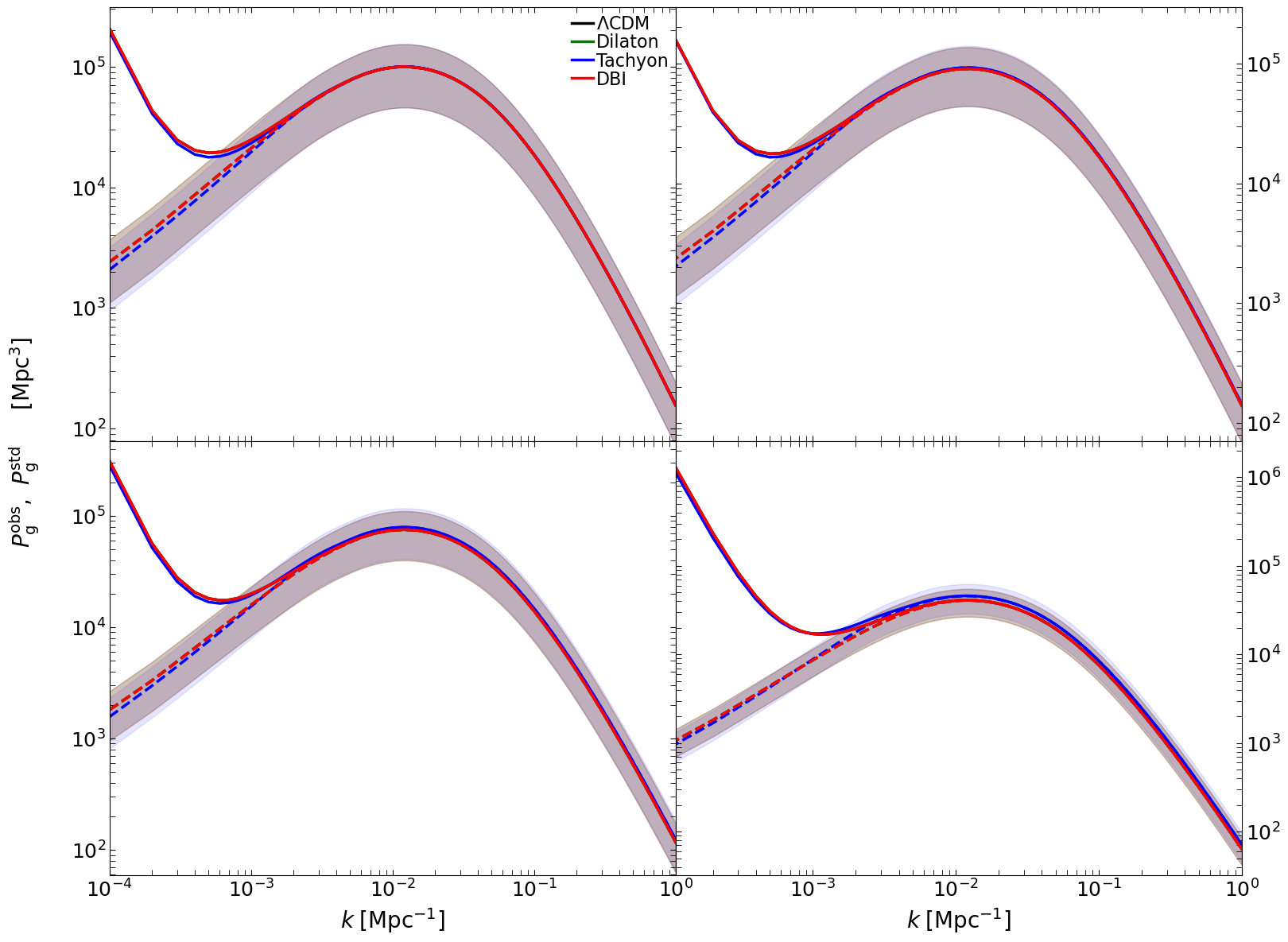} 
	\caption{The plots of the standard linear galaxy power spectrum \eqref{Pofk} (dashed lines) and the relativistic linear galaxy power spectrum \eqref{Pofk-obs} (solid lines), with respect to wavenumber $k$ along the line of sight ($\mu \,{=}\, 1$), at source redshifts $z_S \,{=}\, 0.1$ (top left), $z_S \,{=}\, 0.5$ (top right), $z_S \,{=}\, 1$ (bottom left), and $z_S \,{=}\, 3$ (bottom right). The shaded regions indicate theoretical variance of the standard linear galaxy power spectra.}
	\label{fig:PofKs}
\end{figure*}

Here we consider the Fourier-space linear power spectrum. As widely done in the literature, we neglect the integrated terms in the overdensity \eqref{Delta_n}--\eqref{relsDelta_n} by assuming the flat-sky (or distant-observer) approximation. Then the standard overdensity \eqref{stdDelta_n} in Fourier space becomes
\begin{align}\label{stdDelta_k}
\Delta^{\rm std}_{\rm g}(k,\mu,z_S) = &\; \Delta_{\rm g}(k,z_S) + \mu^2 \dfrac{k^2}{{\cal H}}\, V_m(k,z_S), \nn
= &\; \left[b(z_S) + \mu^2 f(k,z_S)\right] \Delta_m(k,z_S),
\end{align}
where $\Delta_{\rm g} \,{=}\, b\Delta_m$ with $b$ being the galaxy bias \cite{Desjacques:2016bnm, Baldauf:2011bh, Bartolo:2010ec, Jeong:2011as, Lopez-Honorez:2011emg, Duniya:2016ibg}, and the matter overdensity $\Delta_m$ as given by \eqref{D_A}; $\mu\,{=}\, {-}{\bf n} \cdot {\bf k}/k$ is the direction cosine of the wavevector ${\bf k}$ with respect to the line of sight with $k \,{=}\, |{\bf k}|$ being the wavenumber, $f$ is the redshift-space-distortions factor \cite{Duniya:2015nva}, which measures the clustering contribution of redshift space distortions to the matter overdensity, owing to the shear of peculiar velocity fields. (See e.g. \cite{Duniya:2015nva} for $f$ in terms of generalised growth functions.) The clustering factor $f$ reduces to the well-known matter growth rate in $\Lambda$CDM or on sub-Hubble scales in dynamical dark energy (and modified gravity). 

The associated power spectrum, is given by
\beq\label{Pofk}
P^{\rm std}_{\rm g}(k,\mu,z_S) = \left[b(z_S) + \mu^2 f(k,z_S)\right]^2 P_m(k,z_S),
\eeq
where $P_m$ is the matter power spectrum (e.g.~\cite{Duniya:2015nva}).

The relativistic term \eqref{relsDelta_n} in the observed galaxy overdensity, is given by
\beq\label{relsDelta_k}
\Delta^{\rm rels}_{\rm g}(k,\mu,z_S) = \left[ \dfrac{{\cal A}(k,z_S)}{x^2} + i\mu \dfrac{{\cal B}(k,z_S)}{x}\right] \Delta_m(k,z_S), 
\eeq
where $x \,{=}\, k/{\cal H}$, and
\begin{align*}
{\cal A} \;=&\; x^2\left(\dfrac{{\cal H}'}{{\cal H}^2} + \dfrac{2}{{\cal H} \bar{r}_S } + \dfrac{d\ln\Phi}{d\ln a} -b_e -1\right) \dfrac{\Phi}{\Delta_m} \nn
&\;\;+\, \left(3 - b_e\right)f,\\
{\cal B} \;=&\; \left(b_e - \dfrac{{\cal H}'}{{\cal H}^2} - \dfrac{2}{{\cal H}\bar{r}_S}\right) f, 
\end{align*}
with $\mu$ and $f$ being as given in \eqref{stdDelta_k} and, ${\cal A}$ and ${\cal B}$ being the potential and the Doppler factors, respectively.

\begin{figure*}\centering
\includegraphics[scale=0.33]{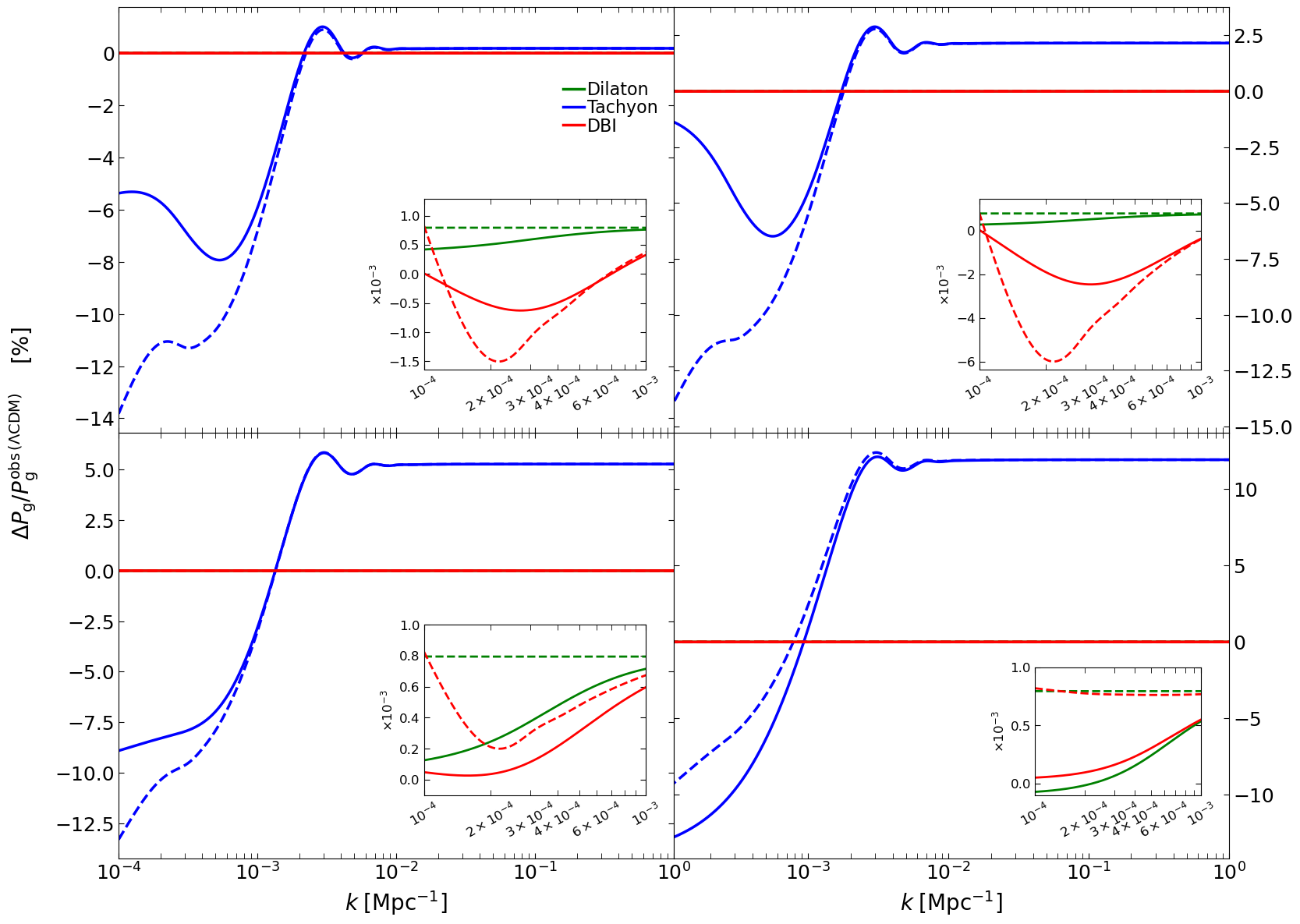} 
\caption{The plots of the percentage deviations of k-essence from the cosmological constant in the relativistic linear galaxy power spectrum $P^{\rm obs}_{\rm g}$ (solid lines) and the standard linear galaxy power spectrum $P^{\rm std}_{\rm g}$ (dashed lines), with respect to $k$: at source redshifts $z_S \,{=}\, 0.1$ (top left), $z_S \,{=}\, 0.5$ (top right), $z_S \,{=}\, 1$ (bottom left), and $z_S \,{=}\, 3$ (bottom right), where $\Delta{P}_{\rm g} \,{=}\, P^{\rm obs}_{\rm g}-P^{\rm obs\, (\Lambda CDM)}_{\rm g}$ for solid lines (and similarly for dashed lines, with $P^{\rm std}_{\rm g}$).}
\label{fig:k2LCDM_PofK_fracs_inset}
\end{figure*}

Thus, the relativistic galaxy linear power spectrum, from \eqref{stdDelta_k} and \eqref{relsDelta_k}, is given by
\beq\label{Pofk-obs}
\dfrac{P^{\rm obs}_{\rm g}}{P_m} = \left(b+\mu^2 f\right)^2 + 2\left(b+\mu^2 f\right) \dfrac{{\cal A}}{x^2} + \dfrac{{\cal A}^2}{x^4} + \mu^2\dfrac{{\cal B}^2}{x^2},
\eeq
where $P_m$ is as given by \eqref{Pofk}, and we use the Euclid fittings given by \cite{Euclid:2019clj, Villa:2017yfg, Duniya:2022miz}
\bea\label{b_g}
b(z) &=& \sqrt{z+1},\\ \label{calN}
{\cal N}(z) &=& 3.5\times 10^8\, z^2 \exp\left[-(z/z_0)^{3/2}\right]
\eea
with ${\cal N}(z)$ being the background number per unit solid angle per redshift, $z_0 \,{=}\, z_{\rm mean}/1.412$ and $z_{\rm mean} \,{=}\, 0.9$, and we compute the evolution bias in ${\cal A}$ and ${\cal B}$, by (e.g.~\cite{Duniya:2022miz})
\beq\label{b_e}
b_e = \dfrac{2a^2}{{\cal H}r} + \dfrac{\partial\ln{\cal N}}{\partial\ln{a}} +\dfrac{{\cal H}'}{{\cal H}^2} - 1.
\eeq
We use \eqref{b_g}--\eqref{b_e} and $\mu \,{=}\, 1$ for all numerical analysis. We assume very large scales where perturbations are linear, and galaxies trace the same trajectory as the underlying matter. (We use adiabatic initial conditions, e.g. \cite{Duniya:2013eta, Duniya:2015nva, Duniya:2015dpa, Duniya:2019mpr}, to solve the perturbations equations.)

As stated in the introduction, the aim in this work is not to fit the k-essence models to the data, but to show how relativistic effects in kCDM differ from those in $\Lambda$CDM, on very large scales. Thus, throughout we refrain from statements on detectability of the relativistic effects or distinguishability between the models. Moreover, we should clarify that our background normalization at today effectively does not affect scale-dependent changes in the power spectra; it only determines whether or not they coincide on small scales (for the different models).

\begin{figure*}\centering
\includegraphics[scale=0.33]{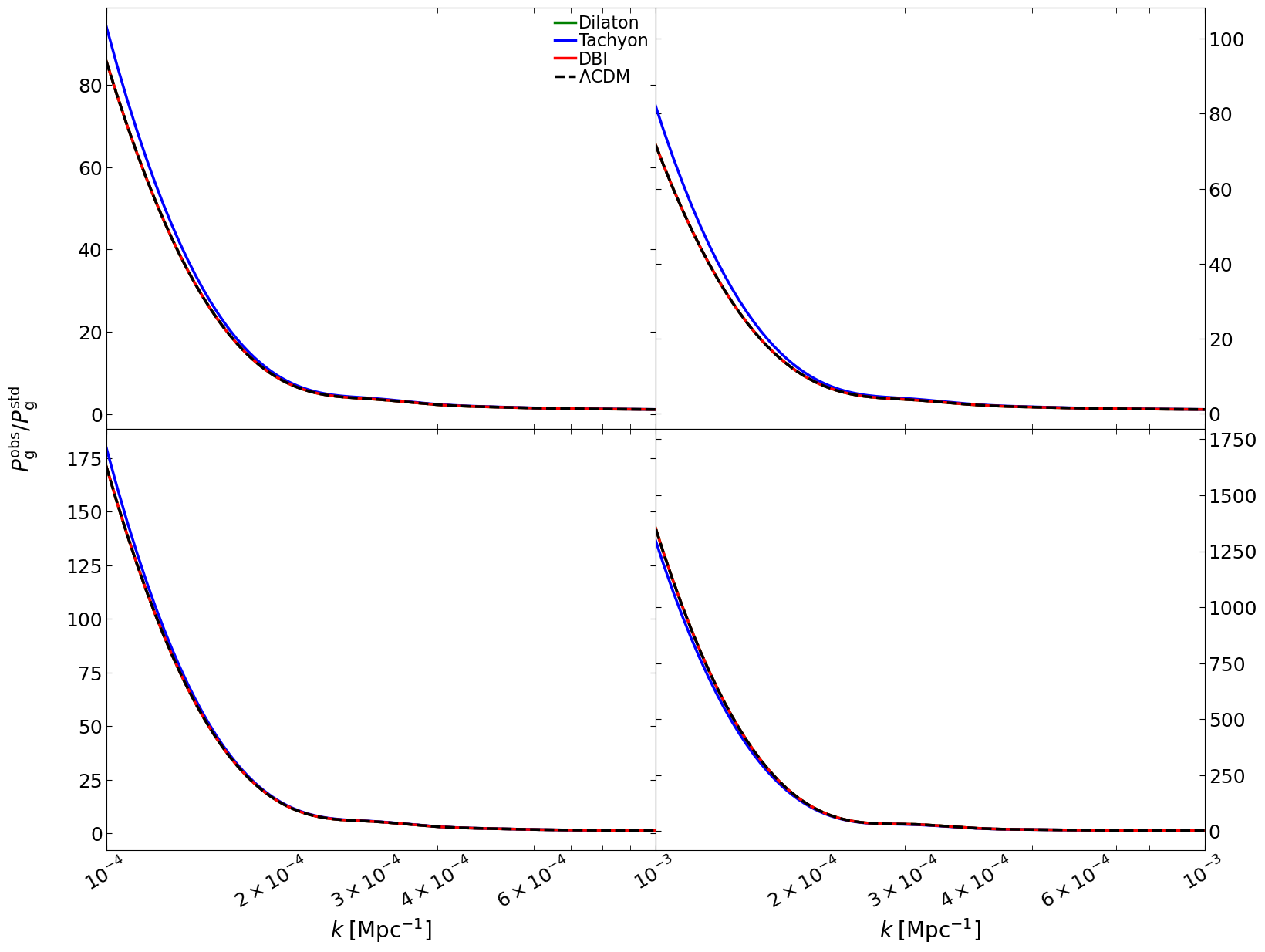} 
\caption{The plots of the ratios of the linear galaxy power spectra---the relativistic to the standard, as in \eqref{Pofk-obs}---with respect to $k$, at source redshifts $z_S \,{=}\, 0.1$ (top left), $z_S \,{=}\, 0.5$ (top right), $z_S \,{=}\, 1$ (bottom left), and $z_S \,{=}\, 3$ (bottom right). The plots show the total effect of the relativistic corrections, relative to the redshift space distortions corrected galaxy clustering overdensity.} 
\label{fig:PofKs_ratios}
\end{figure*}

In Fig.~\ref{fig:PofKs} we show the plots of the radial standard linear galaxy power spectrum \eqref{Pofk}, $P^{\rm std}_{\rm g}(k,z_S)$, and relativistic linear galaxy power spectrum \eqref{Pofk-obs}, $P^{\rm obs}_{\rm g}(k,z_S)$, at source redshifts $z_S \,{=}\, 0.1,\, 0.5,\, 1.0$ and $3.0$, accordingly, for k-essence and the cosmological constant. We also indicate theoretical variance of $P^{\rm std}_{\rm g}(k,z_S)$ as shaded regions. We see that at all the $z_S$ values the relativistic galaxy power spectrum diverges from the standard power spectrum on very large scales. This is caused by the relativistic corrections~\eqref{relsDelta_k} in the observed galaxy number overdensity, as they are known to give such effects (see e.g.~\cite{Jeong:2011as, Duniya:2015nva}). We also see that, although the amplitude of $P^{\rm std}_{\rm g}(k,z_S)$ remains nearly the same for all models at all $z_S \,{\leq}\, 3$, that of $P^{\rm obs}_{\rm g}(k,z_S)$ grows with $z_S$, particularly $z_S \,{\geq}\, 1$, on scales $k \,{\lesssim}\, 10^{-3}\, \mathrm{Mpc}^{-1}$; whereas, on scales $k \,{>}\, 10^{-3}\, \mathrm{Mpc}^{-1}$ the amplitude of $P^{\rm obs}_{\rm g}(k,z_S)$ matches that of $P^{\rm std}_{\rm g}(k,z_S)$. This is consistent with known effects of relativistic corrections: they grow on large scales and at high redshifts. This can be understood from \eqref{Pofk-obs}: the relativistic terms scale as ${\sim}\, ({\cal H}/k)^n$ with $n \,{\geq}\, 2$ so that on small scales they become negligible, and
\beq\nonumber
P^{\rm obs}_{\rm g}/P_m \sim (b+\mu^2 f)^2 = P^{\rm std}_{\rm g}/P_m,
\eeq
and the relativistic linear galaxy power spectra match the standard linear galaxy power spectra. On large scales, the relativistic terms become the leading terms and grow with decreasing $k$. Moreover, effects of the relativistic corrections here do not seem to reveal any noticeable deviations or differences among the dark energy models in the linear galaxy power spectra. This is understandable since at leading regime, these effects are driven by $({\cal H}/k)^n$, which is about the same for all the models, at the given redshifts.

In Fig.~\ref{fig:k2LCDM_PofK_fracs_inset} we show the plots of the percentage deviations of kCDM models from $\Lambda$CDM in both the relativistic and the standard linear galaxy power spectra, $P^{\rm obs}_{\rm g}(k,z_S)$ and $P^{\rm std}_{\rm g}(k,z_S)$, respectively, at source redshifts $z_S \,{=}\, 0.1,\, 0.5,\, 1.0$ and $3.0$. As expected, from Figs.~\ref{fig:wx+cs2} and \ref{fig:PofKs}, the dilaton and the DBI field are identical with the cosmological constant (with deviations at ${\lesssim}\, 0.006\%$) in the linear galaxy power spectra, at all the source redshifts. Moreover, we see that there is a constant deviation between the kCDM models and $\Lambda$CDM on scales $k \,{\gtrsim}\, 10^{-2}\, \mathrm{Mpc}^{-1}$, with the deviation between the tachyon and the rest of the models increasing with source redshift  (which by Fig.~\ref{fig:wx+cs2} will persist at $z_S \,{\lesssim}\, 20$). The apparent convergence of the models on the smaller scales at low source redshifts ($z_S \,{\leq}\, 0.1$) is a consequence of our setup; with all models expected to fully coincide on small scales, at $z_S \,{=}\, 0$. On the largest scales ($k \,{<}\, 10^{-2}\, \mathrm{Mpc}^{-1}$), the tachyon shows a significant deviation from the cosmological constant at all the source redshifts. However, on these scales the deviation is a suppression, with the tachyon having less power relative to the cosmological constant. Basically, on larger scales we have power suppression while on smaller scales we have power enhancement. We can try to understand this from the perturbations, since the tachyon can cluster. The relativistic Poisson equation \eqref{Poisson} shows that the clustering of the tachyon on large scales will enhance the gravitational potential for a given matter overdensity, so that to sustain a given magnitude of gravitational potential, the system will require less clustering of matter. Particularly, on larger scales ($k \,{\ll}\, {\cal H}$), the term $\propto (1+w_\phi)(V_\phi-V_m)$ will counter balance the dissipative effect of the friction term ${\propto}\, w_\phi{\cal H}\Delta_\phi$ in \eqref{dDxdt}, and thereby allowing the tachyon to cluster. Note that the negative of this counter-balancing term appears in \eqref{dDmdt} and hence will have an opposite effect in $\Delta_m$. Thus, simultaneously reducing matter clustering in the gravitational potential. On smaller scales, the given counter-balancing term becomes negligible ($k \,{\gg}\, {\cal H}$). In this case we have a damped (or decaying) evolution for $\Delta_\phi$; consequently, the system will require more matter clustering to maintain the magnitude of the gravitational potential. Thus, a universe with the cosmological constant will have more large-scale matter clustering than that with the tachyon, and conversely, relatively less small-scale matter clustering ($w_\phi \,{>}\, {-}1$ for the tachyon, at $z \,{<}\, 20$). 

Moreover, we see that there is more large-scale power suppression in $P^{\rm std}_{\rm g}(k,z_S{<}3)$ than in $P^{\rm obs}_{\rm g}(k,z_S{<}3)$ for the tachyon relative to the cosmological constant, and conversely, more power suppression in $P^{\rm obs}_{\rm g}(k,z_S{=}3)$ than in $P^{\rm std}_{\rm g}(k,z_S{=}3)$. On the other hand, we see that there is equal (constant) increasing small-scale power enhancement in both $P^{\rm obs}_{\rm g}(k,z_S)$ and $P^{\rm std}_{\rm g}(k,z_S)$ at all $z_S$. Thus, this implies that neglecting relativistic corrections can lead to the overestimation of the deviation of the tachyon from the cosmological constant in the linear galaxy power spectrum, at lower source redshifts ($z_S \,{<}\, 3$), and conversely at higher source redshifts ($z_S \,{\geq}\, 3$), the underestimation of the deviation on the same scales. This can introduce a scale- and redshift-dependent bias that prevent identifying the true relative imprint of the tachyon and hence the ability of properly distinguishing it from the cosmological constant in the cosmological parameters.

In Fig.~\ref{fig:PofKs_ratios} we give the plots of the ratio of the linear galaxy power spectra, $P^{\rm obs}_{\rm g}(k,z_S)/P^{\rm std}_{\rm g}(k,z_S)$, for each model at source redshifts $z_S \,{=}\, 0.1,\, 0.5,\, 1.0$ and $3.0$. This ratio measures the total effect of the relativistic corrections \eqref{relsDelta_k} in the observed galaxy overdensity relative to the (standard) redshift-space-distortions corrected overdensity \eqref{stdDelta_k}. We see that, although there is quite significant growth in the amplitude of the total relativistic effect on very large scales, there is no substantial deviation among the models; only the tachyon shows promise of depature at low source redshifts ($z_S \,{\leq}\, 0.5$). The total relativistic effect for the cosmological constant is identical with that for the dilaton and the DBI models, respectively. This suggests that the given kCDM models will be degenerate with one another and with $\Lambda$CDM in the cosmological parameters. This degeneracy is largely determined by the given background dynamics (Fig.~\ref{fig:wx+cs2}, left panel), which follows from our choice of parameters. For the dilaton and the DBI field, we have $(1+w_\phi) \,{\ll}\, 1$, which cancels out their effect in the gravitational and the matter perturbations equations, making them have the same $\Phi$, $\Delta_m$, and $V_m$ evolutions, respectively, as $\Lambda$CDM (see \S\ref{subsec:Perturbations}). However, for the tachyon $(1+w_\phi) \sim 0.2$ and is no longer negligible, but since its perturbations enter the equations as cofactors of $(1+w_\phi)\Omega_\phi$, which is small,  its effect is diminished proportionally. Thus, all models (but more for the dilaton and the DBI field) will have nearly the same perturbations or their ratios e.g. $f(k,z)$ and $\Phi(k,z)/\Delta_m(k,z)$ where $P_m(k,z) \sim |\Delta_m(k,z)/\Phi(k,0)|^2$. Note that in Fig.~\ref{fig:k2LCDM_PofK_fracs_inset}, the actual background dynamics factors into the ratios at $z_S \,{>}\, 0$; whereas here, each model has identically the same background for both $P^{\rm obs}_{\rm g}(k,z_S)$ and $P^{\rm std}_{\rm g}(k,z_S)$ (hence nullifying the background effect).

In general, our background normalization at today allowed for a clear study of how relativistic corrections in number counts of galaxy redshift surveys respond to k-essence clustering.  In the background, the different k-essence models exhibit different dynamical behaviours despite converging to similar late-time cosmologies. The dilaton and the DBI fields both assume ``freezing'' dynamics: they are either tracking the dominant matter component at early times (the dilaton) or are effectively frozen early (the DBI field), before rapidly converging to the cosmological constant. This behaviour is driven by strong Hubble friction, which suppresses the field evolution and enforces a quasi-static regime. On the other hand, the tachyon traces a ``thawing'' trajectory: it remains frozen with $w_\phi \,{\simeq}\, {-}1$ at early times, and becomes dynamical at later epochs ($a \,{\gtrsim}\, 10^{-2}$). These background evolutions directly influence the clustering properties of the fields through their sound speeds \eqref{dil-cs2-w}--\eqref{dbi-cs2-w}: the dilaton has a vanishing sound speed at late times, the tachyon has $c^2_{s\phi} \,{=}\, {-}w_\phi$, and the DBI field remains luminal, $c^2_{s\phi} \,{\simeq}\, 1$. Consequently, while the dilaton and the DBI field behave as nearly homogeneous dark energy on observable scales, the tachyon retains the capacity to cluster on large scales. 

\begin{figure*}\centering
\includegraphics[scale=0.33]{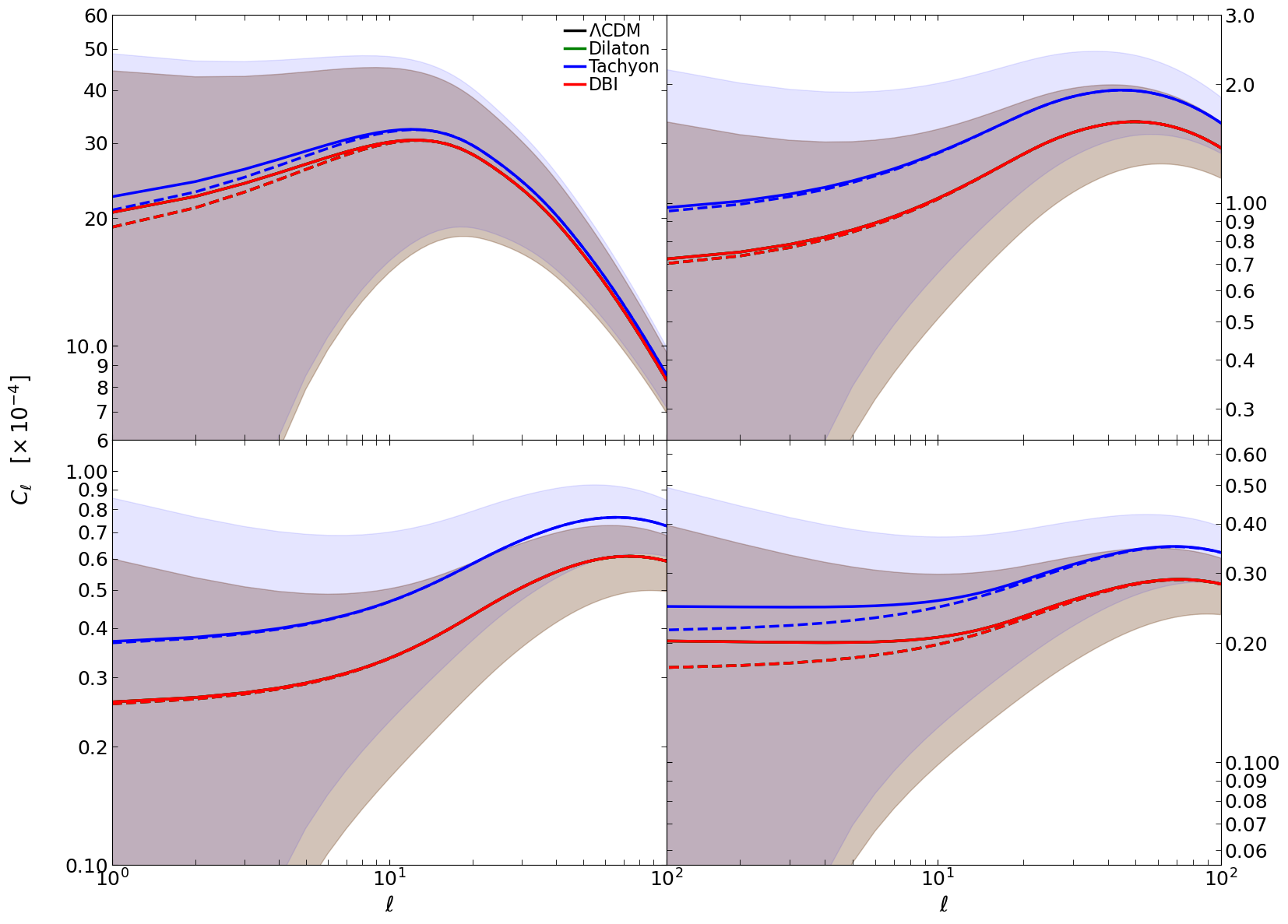} 
\caption{The plots of the relativistic angular galaxy power spectrum \eqref{Cls} (solid lines) and the standard angular galaxy power spectrum (dashed lines), which is computed with $f^{\rm std}_\ell$ as given in \eqref{f_ell}, with respect to multipoles $\ell$. The angular power spectra are given at source redshifts $z_S \,{=}\, 0.1$ (top left), $z_S \,{=}\, 0.5$ (top right), $z_S \,{=}\, 1$ (bottom left), and $z_S \,{=}\, 3$ (bottom right). The shaded regions indicate cosmic variance with sky fraction $f_{\rm sky} \,{=}\, 0.375$ (consistent with Euclid~\cite{Euclid:2019clj}).}
\label{fig:stdtotCls}
\end{figure*}

This distinction becomes critical at the perturbative level. Figs.~\ref{fig:PofKs}--\ref{fig:PofKs_ratios} show that, for the dilaton and the DBI models, the smallness of $1\,{+}\,w_\phi$ suppresses their contribution to both gravitational and matter perturbation equations, rendering their evolution effectively indistinguishable from the cosmological constant. In contrast, the tachyon with $1\,{+}\,w_\phi \sim 0.2$ over a wide redshift range, contributes non-negligibly to the perturbation dynamics, albeit still modulated by the factor $(1\,{+}\,w_\phi)\Omega_\phi$. This leads to scale-dependent modifications in matter clustering: enhanced dark energy clustering on large scales reduces the required matter overdensity to sustain a given gravitational potential, producing a suppression of power at $k \,{\ll}\, \mathcal{H}$, while the opposite effect occurs on small scales.


\section{The Angular Power Spectrum}
\label{sec:Cls}

The angular power spectrum observed at a source redshift $z_S$, is given by
\beq\label{Cls} 
C_\ell(z_S) = \dfrac{4}{\pi^2} \left(\dfrac{9}{10}\right)^2 \int{dk\, k^2 T(k)^2 P_{\Phi_p}(k) \Big|f_\ell(k,z_S) \Big|^2 },
\eeq
where $T(k)$ is the linear transfer function, $P_{\Phi_p}$ is the power spectrum of primordial gravitational potential $\Phi_p$;
\begin{align}\label{f_ell_defn}
f_\ell(k,z_S) =&\; f^{\rm std}_\ell(k,z_S) + f^{\rm rels}_\ell(k,z_S),\\ \label{f_ell}
=&\; b \check{\Delta}_m(k,z_S) j_\ell(kr_S) -\dfrac{\partial_r \check{V}^\parallel_m(k,z_S)}{{\cal H}} \partial^2_{kr} j_\ell(kr_S) \nn
& +\; \dfrac{2}{r_S}\int^{r_S}_0{dr \dfrac{\left(r_S - r\right)}{r} \ell\left(1+\ell\right) j_\ell(kr)\, \check{\Phi}(k,r) } \nn
& +\; f^{\rm rels}_\ell(k,z_S) ,
\end{align}
where $b(z_S)$ is as in \eqref{b_g}, $\check{\Delta}_m \,{=}\, \Delta_m/\Phi_d$ (similarly for $\check{V}_m$ and $\check{\Phi}$) with $\Phi_d$ being the gravitational potential at decoupling (see e.g.~\cite{Duniya:2013eta, Duniya:2015nva, Duniya:2019mpr, Duniya:2022vdi, Duniya:2023xgx}), and 
\begin{align}\label{f_ell_rels}
f^{\rm rels}_\ell(k,z_S) =&\; 2 \left(b_e - \dfrac{{\cal H}'}{{\cal H}^2} - \dfrac{2}{r_S {\cal H}}\right) \int^{r_S}_0{dr\, j_\ell(kr) \check{\Phi}'(k,r) }\nn
&+ \left(b_e - \dfrac{{\cal H}'}{{\cal H}^2} - \dfrac{2}{r_S {\cal H}}\right) \check{V}^\parallel_m(k,z_S) \partial_{kr} j_\ell(kr_S) \nn
&+ \left(\dfrac{{\cal H}'}{{\cal H}^2} + \dfrac{2}{r_S {\cal H}} - b_e - 1\right) \check{\Phi}(k,z) j_\ell(kr) \nn
&+\; \left(3 - b_e\right){\cal H}\check{V}_m(k,z_S) j_\ell(kr_S) \nn
&+\; \dfrac{4}{r_S}\int^{r_S}_0{dr\, j_\ell(kr)\, \check{\Phi}(k,r)} \nn
&+\; j_\ell(kr_S)\dfrac{1}{{\cal H}}\check{\Phi}'(k,z_S)  ,
\end{align}
where $b_e(z_S)$ is as in \eqref{b_e}, $j_\ell$ is the spherical Bessel function, we denote $\partial_{kr} \,{=}\, \partial / \partial(kr)$ and $V_\parallel \,{=}\, {-}{\bf n} \,{\cdot}\, {\bf V} \,{=}\, \partial_rV$, which gives the velocity component along the line of sight with $V$ being the (gauge-invariant) velocity potential.

In Fig.~\ref{fig:stdtotCls} we show the plots of the relativistic angular galaxy power spectrum \eqref{Cls}, $C_\ell(z_S)$, and the standard angular galaxy power spectrum, $C^{\rm std}_\ell(z_S)$, as computed with $f^{\rm std}_\ell$ in \eqref{f_ell}. We give $C_\ell(z_S)$ and $C^{\rm std}_\ell(z_S)$ at source redshifts $z_S \,{=}\, 0.1,\, 0.5,\, 1.0$ and $3.0$ (as in Figs.~\ref{fig:PofKs}-\ref{fig:PofKs_ratios}). Moreover, we indicate cosmic variance as shaded regions, with sky fraction $f_{\rm sky} \,{=}\, 0.375$~\cite{Euclid:2019clj, Villa:2017yfg, Duniya:2022miz}. We see that both $C_\ell(z_S)$ and $C^{\rm std}_\ell(z_S)$ for the dilaton and the DBI field appear to coincide with those of the cosmological constant on all scales $\ell$, at all $z_S$; whereas for the tachyon, both $C_\ell(z_S)$ and $C^{\rm std}_\ell(z_S)$ give noticeable separation from those of the other models. These behaviours can be understood from previous discussions (see \S\ref{subsec:wx-cs2} and \S\ref{sec:Pks}). Nevertheless, it is worthy of note that the tachyon shows significantly larger deviations from the other models in $C_\ell(z_S)$ and $C^{\rm std}_\ell(z_S)$ than in $P^{\rm obs}_{\rm g}(k,z_S)$ and $P^{\rm std}_{\rm g}(k,z_S)$, at the same $z_S$ (see Fig.~\ref{fig:PofKs}). Also, the given deviations are positive (growth) on all $\ell$ and at all $z_S$, and appear to grow with $z_S$. As previously stated, the relative growth in angular power for the tachyon with increase in $z_S$ can be expected from our background normalization of the models at $z \,{=}\, 0$: the power spectra will coincide on small scales (larger $\ell$) at $z \,{=}\, 0$, but will deviate at $z \,{>}\, 0$. 

\begin{figure*}\centering
\includegraphics[scale=0.33]{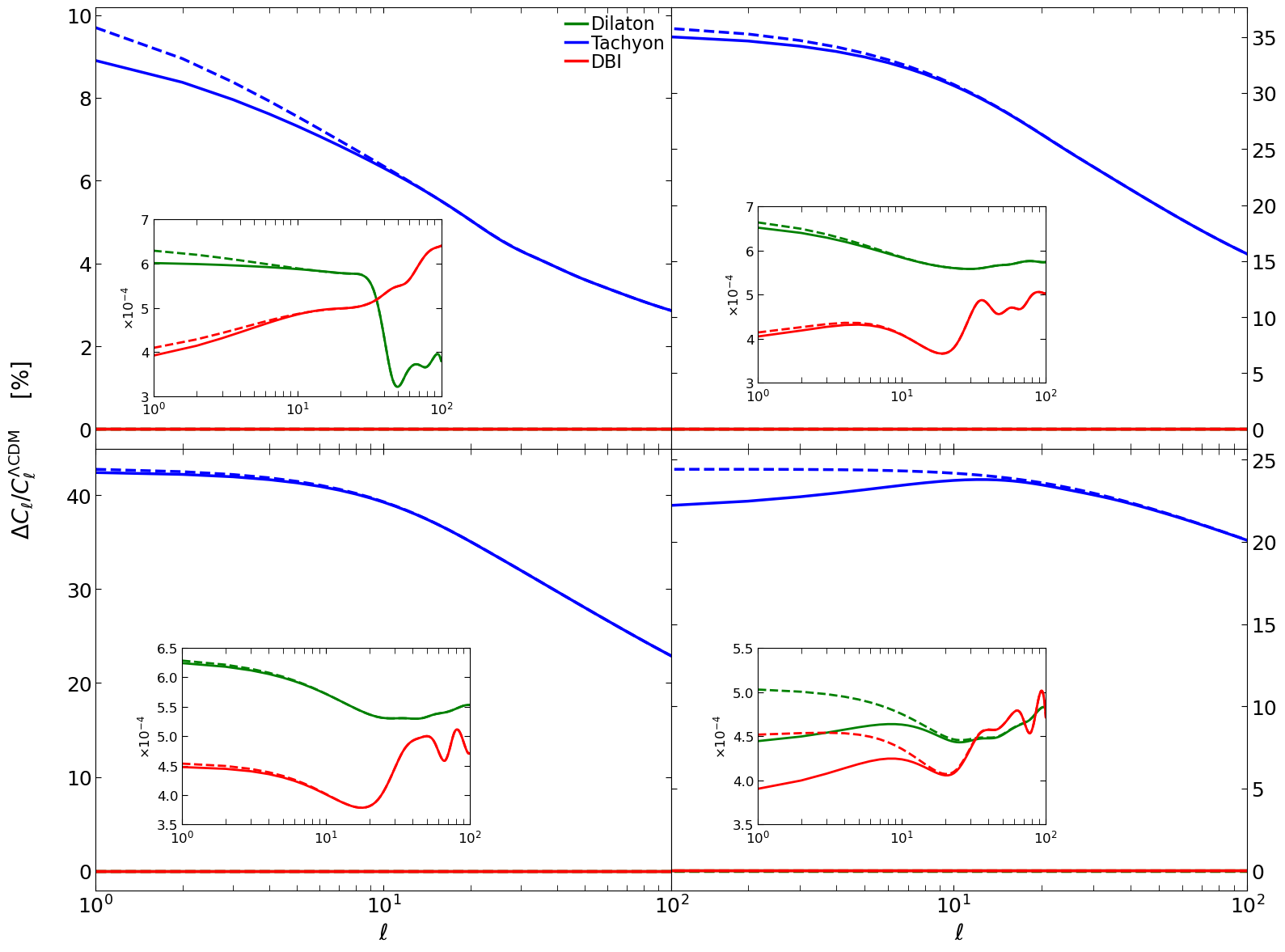} 
\caption{The plots of the fractional deviations (in percentage) of k-essence from the cosmological constant in the relativistic angular galaxy power spectrum $C_\ell$ (solid lines) and the standard angular galaxy power spectrum $C^{\rm std}_\ell$ (dashed lines), with respect to multipole $\ell$: at source redshifts $z_S \,{=}\, 0.1$ (top left), $z_S \,{=}\, 0.5$ (top right), $z_S \,{=}\, 1$ (bottom left), and $z_S \,{=}\, 3$ (bottom right), where $\Delta{C}_\ell \,{=}\, C_\ell-C^{\Lambda\rm CDM}_\ell$ for solid lines (and similarly for dashed lines, with $C^{\rm std}_\ell$).}
\label{fig:k2LCDM_Cls_fracs}
\end{figure*}

Moreover, we see that although the largest deviation of the tachyon from the other models in both $C_\ell(z_S)$ and $C^{\rm std}_\ell(z_S)$ occurs at $z_S \,{=}\, 0.5$ and $z_S \,{=}\, 1.0$, we get the smallest deviations between $C_\ell(z_S)$ and $C^{\rm std}_\ell(z_S)$ for each of the models at the same $z_S$. This suggests that the total effect of the relativistic corrections in the angular galaxy power spectrum becomes negligible at the given $z_S$, and hence the observed angular power spectrum can suitably be approximated by the standard angular power spectrum to describe the true angular distribution of galaxies, for each model at these epochs. On the other hand, at $z_S \,{=}\, 0.1$ and $z_S \,{=}\, 3.0$, $C_\ell(z_S)$ shows relatively substantial deviations from $C^{\rm std}_\ell(z_S)$ for each model; thus, the standard angular power spectrum can no longer be used to characterise the true galaxy distribution in any of the models. Moreover, in contrast, the fact that we do not see similar size of deviations in $P^{\rm obs}_{\rm g}(k,z_S)$ and $P^{\rm std}_{\rm g}(k,z_S)$ can be attributed to the fact that we applied the flat-sky approximation for the linear galaxy power spectrum (as widely done in the literature), and neglected integral corrections (weak-lensing, ISW, and time-delay terms); whereas, multipole expansion allows for natural inclusion of these terms in the angular galaxy power spectrum. Although the integral terms may be individually negligible at $z_S \,{\lesssim}\, 1$, their cross-term with the density, Doppler or redshift space distortions terms can be significant, and hence give significant contribution (or deviation). Furthermore, the apparent better sensitivity to relativistic corrections by $C_\ell(z_S)$ compared to $P^{\rm obs}_{\rm g}(k,z_S)$ (see Figs.~\ref{fig:PofKs} and~\ref{fig:stdtotCls}) tends to position the angular galaxy power spectrum as a better cosmological probe of relativistic effects over the linear galaxy power spectrum.

\begin{figure*}\centering
\includegraphics[scale=0.33]{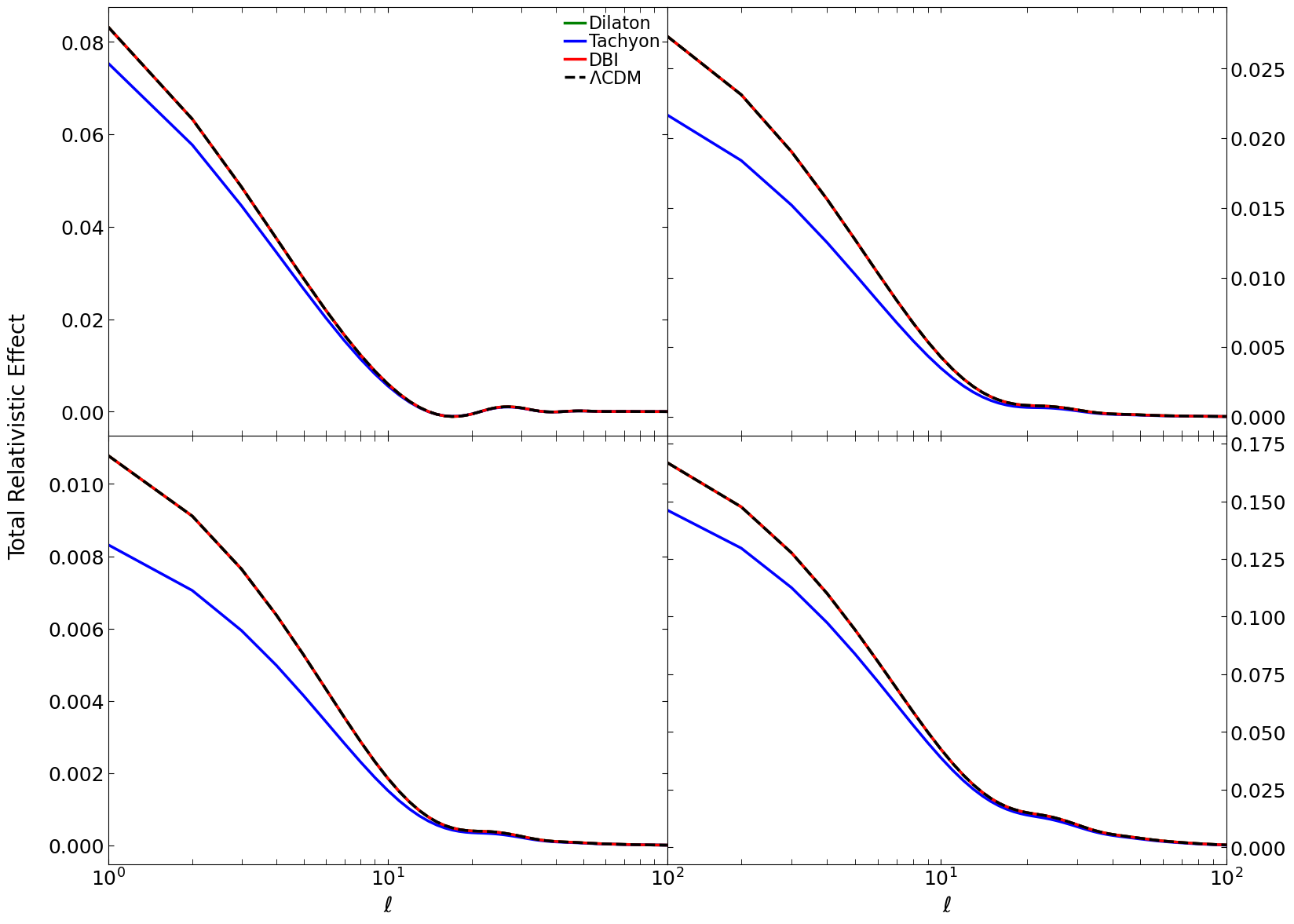} 
\caption{The plots of the total relativistic effect in the angular galaxy power spectrum in kCDM and $\Lambda$CDM, with respect to multipole $\ell$, at source redshifts $z_S \,{=}\, 0.1$ (top left), $z_S \,{=}\, 0.5$ (top right), $z_S \,{=}\, 1$ (bottom left), and $z_S \,{=}\, 3$ (bottom right), where the total relativistic effect ${\cal R}_\ell \,{=}\, C_\ell/C^{\rm std}_\ell -1$. All notations are as in Fig.~\ref{fig:stdtotCls}.}
\label{fig:total_rels_effects}
\end{figure*}

In Fig.~\ref{fig:k2LCDM_Cls_fracs} we give the plots of the percentage deviations of k-essence from the cosmological constant in both $C_\ell(z_S)$ and $C^{\rm std}_\ell(z_S)$, with respect to multipole $\ell$, at source redshifts $z_S \,{=}\, 0.1,\, 0.5,\, 1.0$ and $3.0$. Similar discussion follows as for Fig.~\ref{fig:stdtotCls}. Moreover, by comparing with results in Fig.~\ref{fig:k2LCDM_PofK_fracs_inset}, we see that the percentage deviations of the dilaton and the DBI field from the cosmological constant, on very large scales, are negligible but differ by ${\cal O}(10^1)$ in the linear and the angular galaxy power spectra, $\sim 10^{-3}$ and $\sim 10^{-4}$, respectively, at all the source redshifts (see insets in Figs.~\ref{fig:k2LCDM_PofK_fracs_inset} and \ref{fig:k2LCDM_Cls_fracs}). Also, unlike in $P^{\rm obs}_{\rm g}(k,z_S)$ and $P^{\rm std}_{\rm g}(k,z_S)$ where on large scales (smaller $k$) there is a scale-dependent deviation between the models, and on small scales (larger $k$) there is a (constant) scale-independent deviation at all $z_S$, here we only have a scale-dependent deviation in $C_\ell(z_S)$ and $C^{\rm std}_\ell(z_S)$, at the same $z_S$. This can be understood as a consequence of the integral nature of the angular galaxy power spectrum, where larger-$k$ effects will sum to give totals of the same amplitude, in a given cosmological model.

Furthermore, on large scales, we see that while the deviation of the dilaton in $P^{\rm std}_{\rm g}(k,z_S)$ remains positive and larger than that in $P^{\rm obs}_{\rm g}(k,z_S)$ at all $z_S$, the deviations of the DBI field and the tachyon in $P^{\rm std}_{\rm g}(k,z_S)$ are negative and lesser than those in $P^{\rm obs}_{\rm g}(k,z_S)$ at $z_S \,{<}\, 1$ and $z_S \,{<}\, 3$, respectively; whereas at $z_S \,{\geq}\, 1$ and $z_S \,{\geq}\, 3$, they become positive and greater that those in $P^{\rm obs}_{\rm g}(k,z_S)$ (but still suppressed relative to $\Lambda$CDM). In contrast, all the kCDM models consistently exhibit positive and larger deviations (enhancement; albeit marginal for the dilaton and the DBI field) from $\Lambda$CDM in $C^{\rm std}_\ell(z_S)$ than in $C_\ell(z_S)$ on scales $\ell \,{\lesssim}\, 10$, at all $z_S$. This implies that neglecting relativistic corrections in the angular galaxy power spectrum will lead to the same effect---overestimation of galaxy clustering---in kCDM models on very large scales; whereas, this will lead to different predictions (underestimation or overestimation) in the linear galaxy power spectrum for different kCDM models. This further adds to the apparent relative suitability of the angular galaxy power spectrum, over the linear galaxy power spectrum, as a more robust cosmological tool for probing large-scale imprints of dark energy and relativistic corrections. 

\begin{figure*}\centering
\includegraphics[scale=0.33]{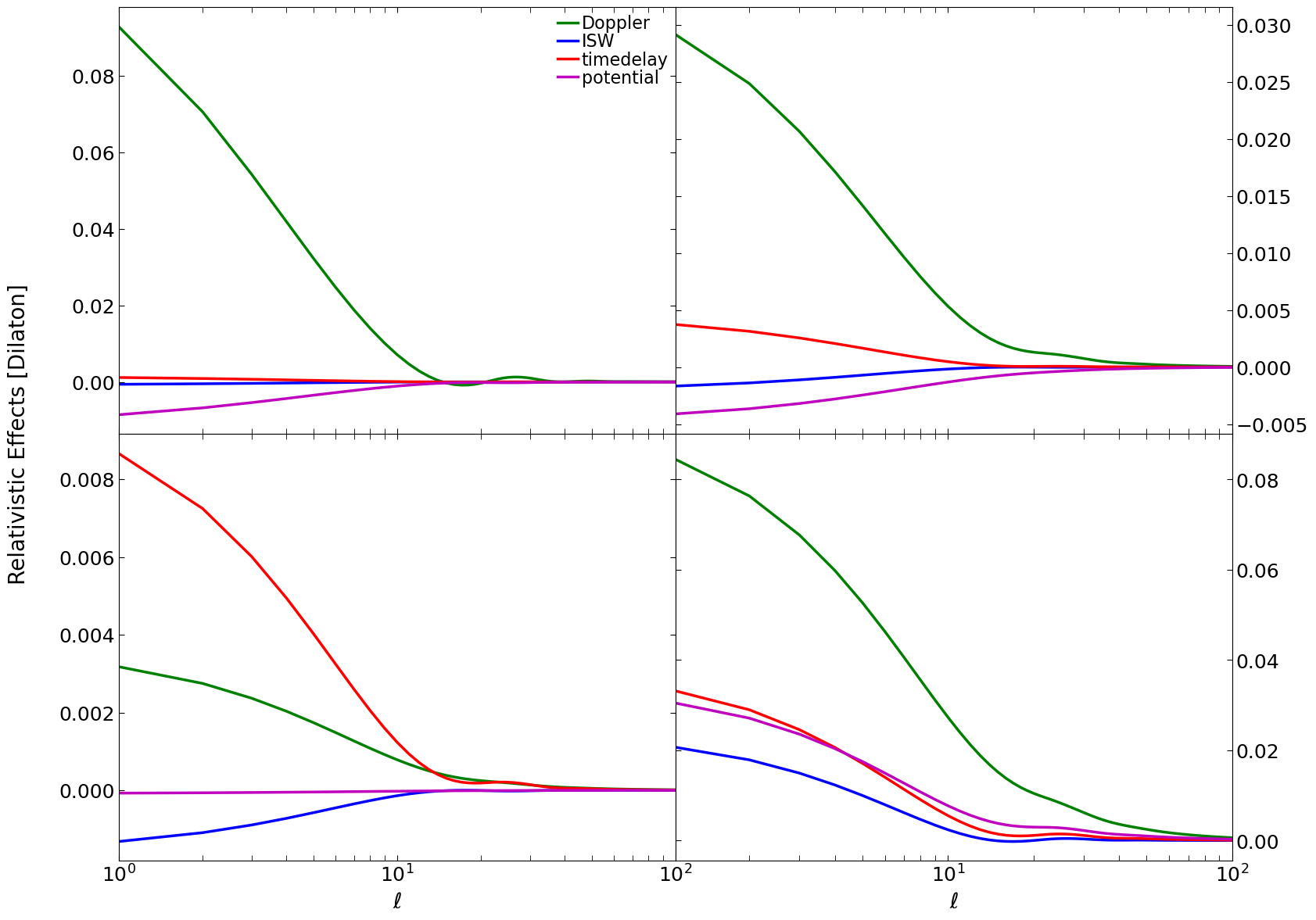} 
\caption{The plots of the individual relativistic effects in the full angular galaxy power spectrum for the dilaton, with respect to multipole $\ell$, at source redshifts $z_S \,{=}\, 0.1$ (top left), $z_S \,{=}\, 0.5$ (top right), $z_S \,{=}\, 1$ (bottom left), and $z_S \,{=}\, 3$ (bottom right). The Doppler effect is measured by ${\cal R}^{\rm Doppler}_\ell = C_\ell/C^{\rm (no\, Doppler)}_\ell -1$, with $C^{\rm (no\, Doppler)}_\ell$ denoting the angular galaxy power spectrum with Doppler correction being neglected; similarly for ISW, time-delay, and (velocity-gravitational) potential corrections, respectively. }
\label{fig:Dilaton_rels_effects}
\end{figure*}

In Fig.~\ref{fig:total_rels_effects} we show plots of the total relativistic effect ${\cal R}_\ell(z_S) = C_\ell(z_S)/C^{\rm std}_\ell(z_S) -1$ in the full, relativistic angular galaxy power spectrum in all the kCDM models and $\Lambda$CDM, with respect to multipole $\ell$, at source redshifts $z_S \,{=}\, 0.1,\, 0.5,\, 1.0$ and $z_S \,{=}\, 3.0$. Consistent with results in Fig.~\ref{fig:PofKs_ratios}, we see the total effect of the relativistic corrections for the dilaton, the DBI field, and the cosmological constant all coinciding in the angular galaxy power spectrum on all scales, at all source redshifts. As previously discussed, this follows from the fact that the given kCDM models have nearly identical background cosmology with $\Lambda$CDM at the given epochs, with $w_\phi(z_S\,{\lesssim}\,100) \simeq {-}1$ (a consequence of our normalization). Similarly, as discussed under Fig.~\ref{fig:PofKs_ratios}, the relative (significant) deviation in background cosmology from the cosmological constant enables the clustering effect of the tachyon to influence the relevant cosmological equations, and thereby resulting in the deviation in $C_\ell(z_S)$. However, the large-scale separation between the tachyon and the cosmological constant in the total relativistic effect in $C_\ell(z_S)$ is larger than in $P^{\rm obs}_{\rm g}(k,z_S)$, at all $z_S$. As discussed under Fig.~\ref{fig:stdtotCls}, these changes can be attributed to integral effects in $C_\ell(z_S)$, which are neglected in $P^{\rm obs}_{\rm g}(k,z_S)$. This suggests that total relativistic effect in the angular galaxy power spectrum holds a relatively better potential than that in the linear galaxy power spectrum (when flat-sky approximation is used) at enhancing deviations between dark energy models that have different background dynamics. Moreover, the total relativistic effect for the tachyon in $P^{\rm obs}_{\rm g}(k,z_S)$ is enhanced relative to that in the other models; whereas in $C_\ell(z_S)$, it is relatively suppressed. But given that for the tachyon we have $w_\phi(z\,{\lesssim}\,20) >  {-}1$ and it clusters on very large scales, it is more sensible to expect large-scale power suppression rather than enhancement. Thus, the relativistic angular galaxy power spectrum appears more realistic. In what follows therefore (for completeness), we focus on individual relativistic effects in the relativistic angular galaxy power spectrum.

In Fig.~\ref{fig:Dilaton_rels_effects} we show, for the dilaton, the plots of the individual relativistic effects in the full angular galaxy power spectrum, with respect to multipole $\ell$, at source redshifts $z_S \,{=}\, 0.1,\, 0.5,\, 1.0$, and $3.0$. Similar results were obtained for the DBI field, the tachyon, and the cosmological constant, and hence specific plots for those models are not given here. We compute the relativistic effect of X at $z_S$ by ${\cal R}^{\rm X}_\ell(z_S) \,{=}\, C_\ell(z_S)/C^{\rm (no\, X)}_\ell(z_S) -1$, where $C^{\rm (no\, X)}_\ell$ denotes the angular galaxy power spectrum with X being neglected: X = Doppler, ISW, time-delay, and (velocity and gravitational) potential corrections, respectively. These results are consistent with previous results by e.g.~\cite{Duniya:2022xcz} (Fig.~2, left panel) for non-interacting phenomenological dark energy, except that the Doppler effect remains the dominant effect at $z_S \,{\leq}\, 1$ for the phenomenological dark energy while here we have the Doppler effect to become subdominant at $z_S \,{=}\, 1$ (but dominant otherwise) for the dilaton. 

We see in Fig.~\ref{fig:Dilaton_rels_effects} that the Doppler effect is the dominant relativistic effect at most $z_S$, except at $z_S \,{=}\, 1$, where it becomes the second dominant effect after the time-delay effect. At the epochs the Doppler effect dominates, it does so by a relatively substantial amount, with respect to the closest subdominant effect. Moreover, its amplitude appears to decrease with increasing source redshift at $z_S \,{\leq}\, 1$, but it still remains the second dominant effect even at its lowest amplitude (with the lowest value being at $z_S \,{=}\, 1$); it becomes dominant again at $z_S \,{=}\, 3$. In general, the Doppler effect stays significant through the epochs. Conversely, the other relativistic effects (ISW, time-delay, and potentials) remain subdominant through most of the epochs, but they show consistent increase in amplitude (growth) with increasing source redshift, attaining their largest amplitude at $z_S \,{=}\, 3$. This large-scale growth in amplitude of the relativistic effects in the full angular galaxy power spectra is already widely known in the literature. In what follows we examine each relativistic effects (separately) in the kCDM models and $\Lambda$CDM.

\begin{figure*}\centering
\includegraphics[scale=0.33]{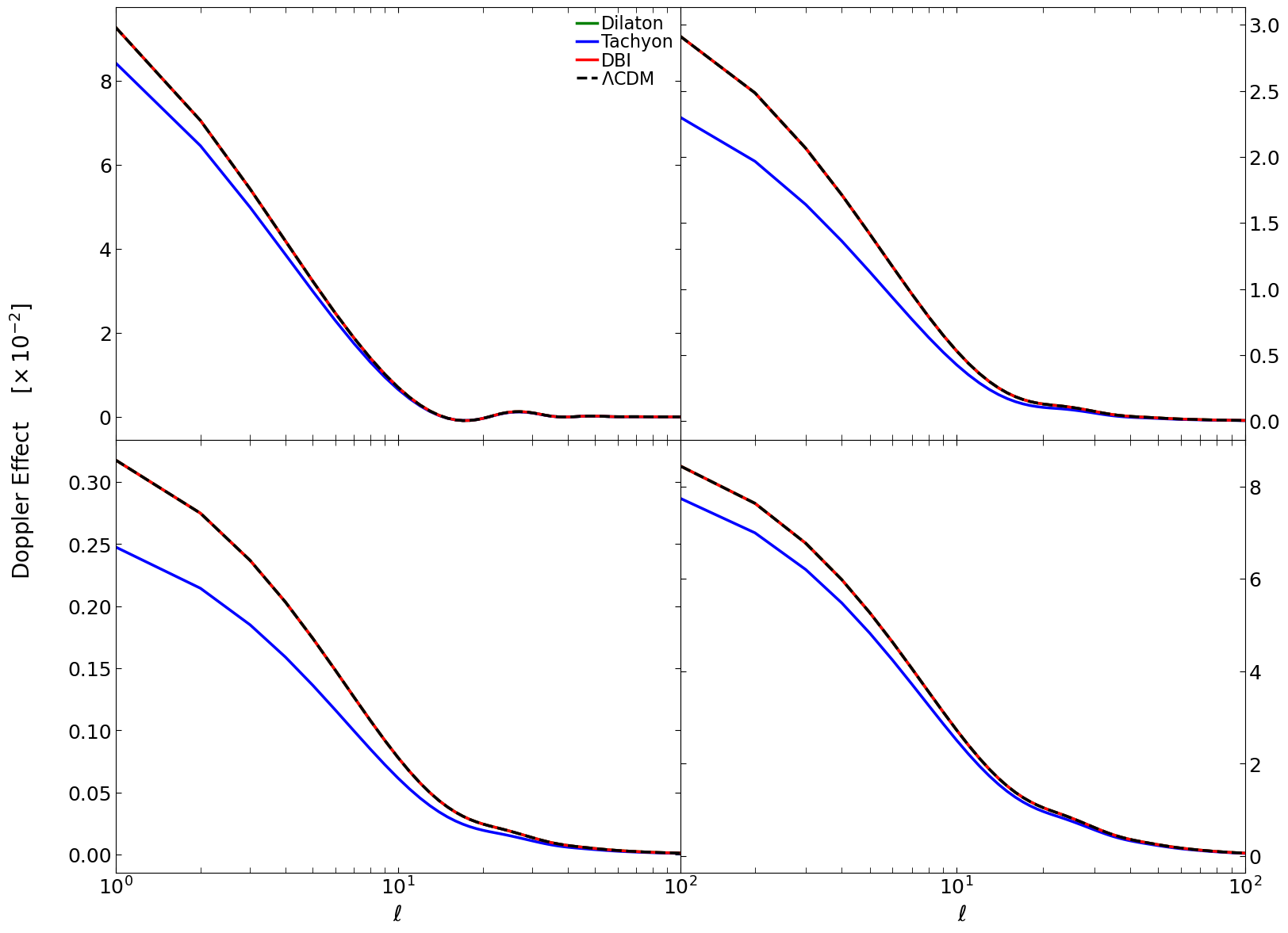} 
\caption{The plots of Doppler effect in the full angular galaxy power spectrum for the kCDM models and $\Lambda$CDM, at source redshifts $z_S \,{=}\, 0.1$ (top left), $z_S \,{=}\, 0.5$ (top right), $z_S \,{=}\, 1$ (bottom left), and $z_S \,{=}\, 3$ (bottom right). All notations are as in Fig.~\ref{fig:Dilaton_rels_effects}. }
\label{fig:Doppler_effect}
\end{figure*}

\begin{figure*}\centering
\includegraphics[scale=0.33]{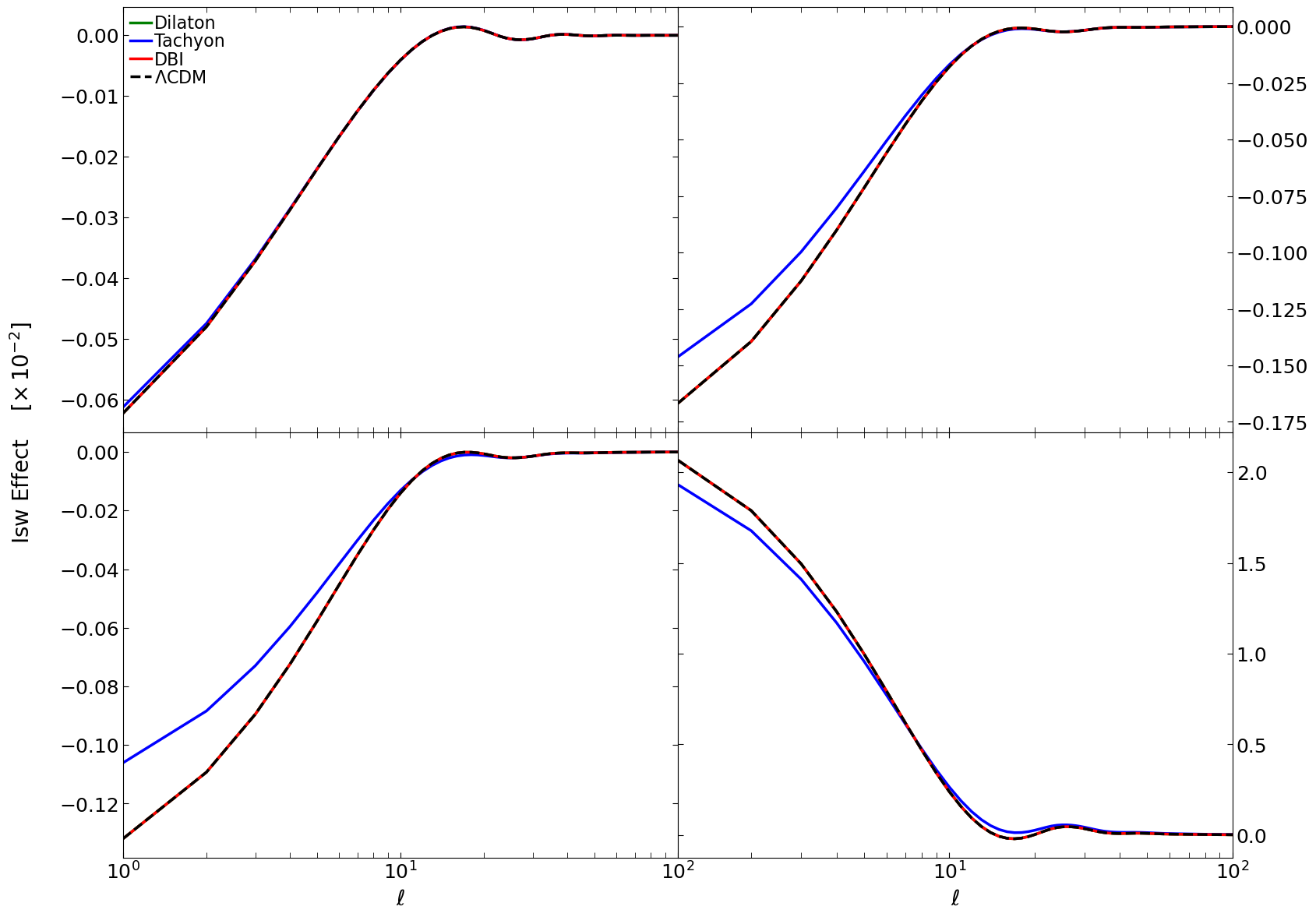} 
\caption{The plots of ISW effect in the full angular galaxy power spectrum for the kCDM models and $\Lambda$CDM, at source redshifts $z_S \,{=}\, 0.1$ (top left), $z_S \,{=}\, 0.5$ (top right), $z_S \,{=}\, 1$ (bottom left), and $z_S \,{=}\, 3$ (bottom right). All notations are as in Fig.~\ref{fig:Dilaton_rels_effects}. }
\label{fig:ISW_effect}
\end{figure*}

\begin{figure*}\centering
\includegraphics[scale=0.33]{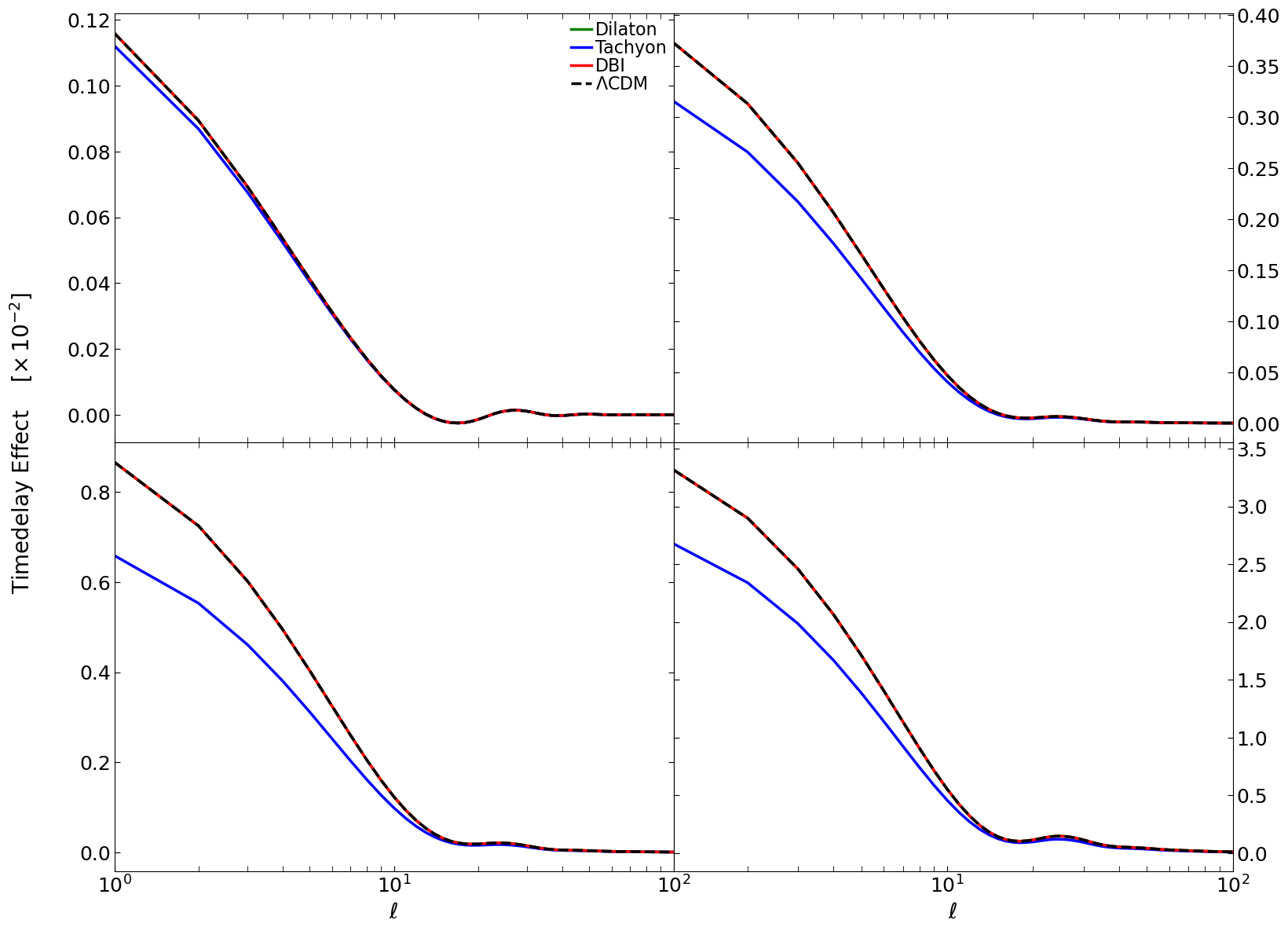} 
\caption{The plots of time-delay effect in the full angular galaxy power spectrum for the kCDM models and $\Lambda$CDM, at source redshifts $z_S \,{=}\, 0.1$ (top left), $z_S \,{=}\, 0.5$ (top right), $z_S \,{=}\, 1$ (bottom left), and $z_S \,{=}\, 3$ (bottom right). All notations are as in Fig.~\ref{fig:Dilaton_rels_effects}. }
\label{fig:timedelay_effect}
\end{figure*}

In Fig.~\ref{fig:Doppler_effect} we show the plots of Doppler effect in the relativistic angular galaxy power spectrum for all the kCDM models and $\Lambda$CDM, at source redshifts $z_S \,{=}\, 0.1,\, 0.5,\, 1.0$, and $3.0$. These plots can reveal any tendency of (or lack of), the Doppler effect to separate dark energy models in the relativistic angular galaxy power spectrum. Consistent with the previous results (Figs.~\ref{fig:PofKs}--\ref{fig:total_rels_effects}) we see that the Doppler effect is identical for the dilaton, the DBI, and the cosmological constant, with amplitude and behaviour as in Fig.~\ref{fig:Dilaton_rels_effects}. However, for the tachyon the Doppler effect is suppressed, relative to that in the other models. This is understandable from previous discussions. The separation between the Doppler effect for the tachyon and the other models suggests that the Doppler effect in the relativistic angular galaxy power spectrum has the potential to distinguish the tachyon from the dilaton, the DBI field, and the cosmological constant on very large scales, at all source redshifts, when they are normalized to the same background cosmology at today. It is also worth noting that the Doppler effect in the relativistic angular galaxy power spectrum behaves similar to the total relativistic effect (Fig.~\ref{fig:total_rels_effects}), with the only difference appearing in amplitude, at different source redshifts: at $z_S \,{<}\, 1$, the Doppler effect has larger amplitude than the total relativistic effect, and at $z_S \,{\geq}\, 1$, the total relativistic effect becomes larger in amplitude than the Doppler effect. This shows that the effect of Doppler correction will be the dominant relativistic effect in the observed angular galaxy power spectrum at low source redshifts ($z_S \,{<}\, 1$), with the other corrections---mostly being proportional to the gravitational potential and its integrals---giving not only subdominant effects but also negative contribution (via their cross terms), which will lower the amplitude of the total relativistic effect at the given source redshifts. Moreover, the fact that the Doppler effect is positive on the largest scales in all the models implies that Doppler correction has a net positive contribution in the full angular galaxy power spectrum: its removal reduces the angular power spectrum, i.e. $C^{\rm (no\, Doppler)}_\ell(z_S) < C_\ell(z_S)$.

\begin{figure*}\centering
\includegraphics[scale=0.33]{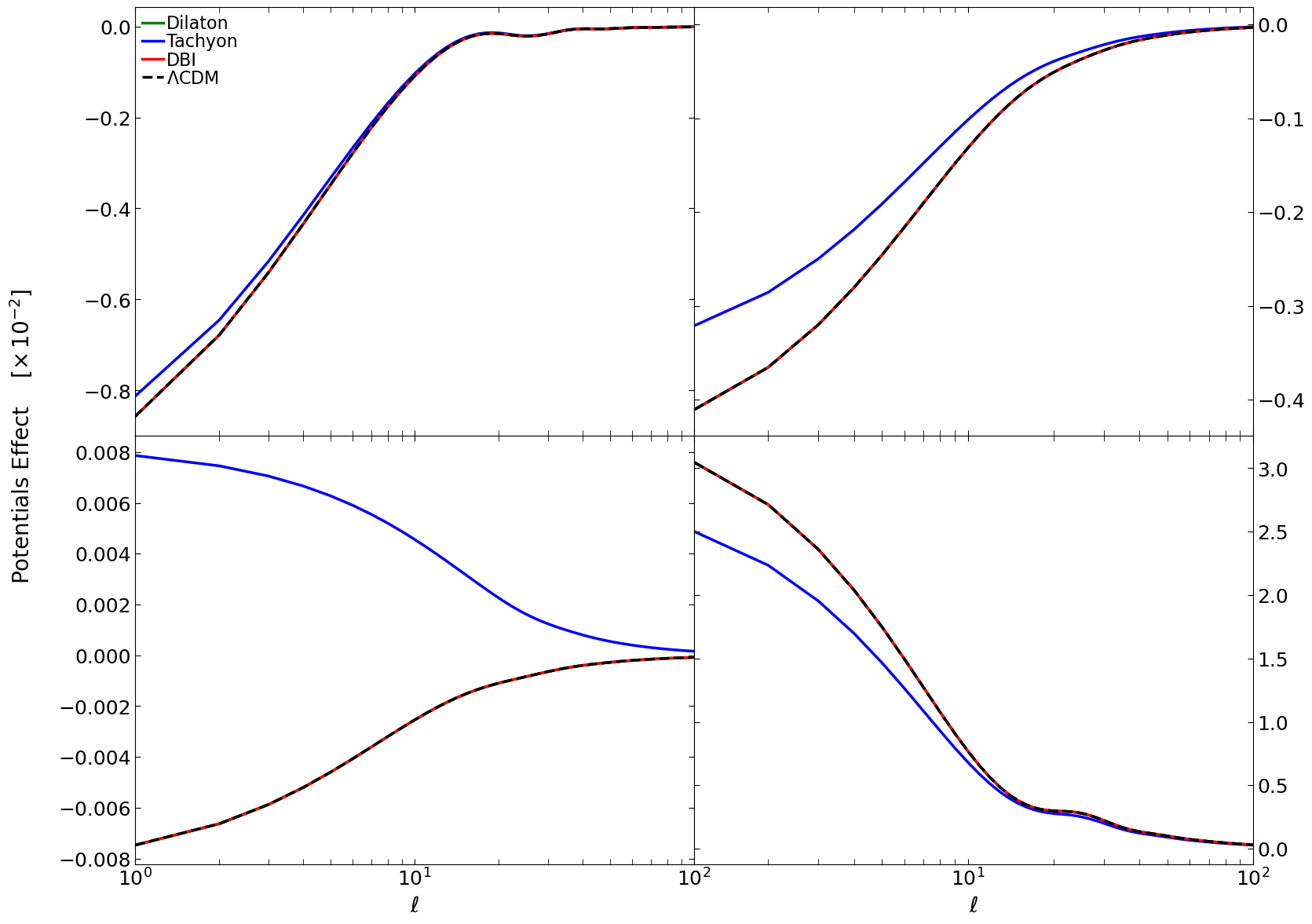} 
\caption{The plots of (velocity and gravitational) potentials effect in the full angular galaxy power spectrum for the kCDM models and $\Lambda$CDM, at source redshifts $z_S \,{=}\, 0.1$ (top left), $z_S \,{=}\, 0.5$ (top right), $z_S \,{=}\, 1$ (bottom left), and $z_S \,{=}\, 3$ (bottom right). All notations are as in Fig.~\ref{fig:Dilaton_rels_effects}. }
\label{fig:potential_effect}
\end{figure*}

In Fig.~\ref{fig:ISW_effect} we show the plots of ISW effect in the relativistic angular galaxy power spectrum for the kCDM models and $\Lambda$CDM, at source redshifts $z_S \,{=}\, 0.1,\, 0.5,\, 1.0$, and $3.0$. The ISW effect is computed as given in Fig.~\ref{fig:Dilaton_rels_effects}. We have that at $z_S \,{<}\, 3$, the amplitude of $C_\ell(z_S)$ is less than that of $C^{\rm (no\, ISW)}_\ell(z_S)$ on the largest scales in all the models, as the ISW effect is having negative amplitudes, at the given $z_S$. In other words, including the ISW correction diminishes the relativistic angular galaxy power spectrum on very large scales in any of the models, at low source redshifts. Thus, at the given epochs, the ISW correction will have net negative effect in the observed angular galaxy power spectrum. Moreover, at these source redshifts, the results suggest that the tachyon will have larger ISW effect in the observed angular galaxy power spectrum than in the other models (if normalized to the same background at today): we see relatively larger amplitude of ISW effect on large scales for the tachyon, which is maximum at $z_S \,{=}\, 1$ and decreases at later epochs---further suggesting that the ISW effect can (in principle) suitably reveal true difference between the tachyon and the other models at $z_S \,{=}\, 1$. Conversely, at $z_S \,{=}\, 3$, the ISW effect becomes positive and grows on very large scales in all the models; with the tachyon now giving the lowest ISW effect. At this source redshift, and possibly higher, the net contribution of the ISW correction will enhance the observed angular galaxy power spectrum, in all the models.

In Fig.~\ref{fig:timedelay_effect} we show the plots of time-delay effect in the relativistic angular galaxy power spectrum for the kCDM models and $\Lambda$CDM, at source redshifts $z_S \,{=}\, 0.1,\, 0.5,\, 1.0$, and $3.0$. The time-delay effect is computed as given in Fig.~\ref{fig:Dilaton_rels_effects}. Unlike the ISW effect, we see the time-delay effect is positive and grows on very large scales for all the models, at all the $z_S$. This implies that the time-delay correction has a net positive contribution in the relativistic angular galaxy power spectrum, at all source redshifts. Moreover, like in the previous results, the time-delay effect for the dilaton, the DBI, and the cosmological constant all coincide on all scales and at all the $z_S$; with the time-delay effect for the tachyon being relatively lower, i.e. the tachyon will act to suppress time-delay effect in the relativistic angular galaxy power spectrum relative to the cosmological constant, at $z_S \,{\leq}\, 3$. Furthermore, the separation between the time-delay effect for the tachyon and that in the other models grows with $z_S$. This suggests that the potential of distinguishing the tachyon from the other models using the time-delay effect in the relativistic angular galaxy power spectrum, in principle, increases with source redshift.

In Fig.~\ref{fig:potential_effect} we show the plots of the combined (velocity and gravitational) potentials effect in the relativistic angular galaxy power spectrum for the kCDM models and $\Lambda$CDM, at source redshifts $z_S \,{=}\, 0.1,\, 0.5,\, 1.0$, and $3.0$. Similarly, the potentials effect is computed as given in Fig.~\ref{fig:Dilaton_rels_effects}. Here we have behaviour as in the ISW effect (Fig.~\ref{fig:ISW_effect}), except that the separation between the potentials effect for the tachyon and that for the rest of the other models is relatively wider; also, at $z_S \,{=}\, 1$ we have the potentials effect for the tachyon become positive on scales $\ell \,{\lesssim}\, 100$. Thus, relative to the ISW effect, the potentials effect will be better at showing the difference between the tachyon and the other models.

In general, across all the $z_S$ values ($ 0.1 \leq z_S \leq 3.0$), the dilaton and the DBI field are indistinguishable from the cosmological constant in both $C_\ell(z_S)$ and $C^{\rm std}_\ell(z_S)$, reflecting their nearly identical background evolution ($w_\phi \,{\simeq}\, {-}1$) under normalization at today. However, the tachyon exhibits positive deviations (enhanced clustering) on all angular scales, with deviations growing with redshift. Notably, these differences are significantly amplified in $C_\ell(z_S)$ compared to $P^{\rm obs}_{\rm g}(k,z_S)$, indicating that angular statistics will be more sensitive to underlying cosmological dynamics. A central result is the non-negligible role of relativistic corrections in $C_\ell$. At intermediate redshifts ($z_S \,{=}\, 0.5,\, 1.0$), $C_\ell(z_S) \approx C^{\rm std}_\ell(z_S)$ so relativistic effects are negligible. However, at low and high redshifts ($z_S \,{=}\, 0.1,\, 3.0$), substantial deviations arise, rendering the standard approximation inadequate. Moreover, the total relativistic effect in $C_\ell(z_S)$ is identical for the dilaton, the DBI field, and the cosmological constant, but significantly different for the tachyon, especially on large scales. Unlike $P^{\rm obs}_{\rm g}(k,z_S)$, where relativistic effects can ``artificially'' enhance tachyon clustering, $C_\ell(z_S)$ yields a physically consistent suppression, aligning with its equation of state ($w_\phi \,{>}\, {-}1$) and clustering properties. Furthermore, the Doppler effect dominates at most redshifts, contributing positively to large-scale power and enabling clear separation of the tachyon model (which shows suppressed Doppler amplitude). The ISW effect is negative at $z_S \,{<}\, 3$ (reducing power) and becomes positive at higher redshifts; it is relatively stronger for the tachyon at intermediate redshifts, aiding discrimination. The time-delay effect is always positive and growing with redshift, but suppressed in the tachyon, with increasing separation at higher redshifts. The potentials effect behaves similar to the ISW effect but provides even stronger model separation, especially for the tachyon.


\section{Conclusion}
\label{sec:Concl}

We presented a systematic investigation of relativistic effects in large-scale galaxy clustering within k-essence cosmologies, focusing on three representative models (dilaton, tachyon, and DBI scalar fields) and contrasting their predictions with those of the concordance $\Lambda$CDM scenario. For the purpose of this work, which is to show how relativistic effects in the k-essence cosmologies differ from those in $\Lambda$CDM
on very large scales (not fitting the models to the data), we normalized all models to share identical present-day background parameters, $\Omega_{m0}$ and $H_0$, thereby isolating the imprints of dark energy dynamics and perturbations on ultra-large scales. 

This approach allowed for a clear, isolated study of how relativistic corrections in galaxy number counts respond to departures from a cosmological constant, independent of late-time background degeneracies.  At the background level, the models exhibited different dynamical behaviours despite converging to similar late-time cosmologies. The dilaton and the DBI fields both realised ``freezing'' dynamics: the dilaton tracked the dominant matter component at early times while the DBI field was effectively frozen already at decoupling, and both rapidly converged to $w_\phi \simeq {-}1$. In contrast, the tachyon followed a ``thawing'' trajectory, remaining frozen with $w_\phi \approx {-}1$ at early times and only evolving dynamically at later epochs ($z \lesssim 100$). These distinct evolutionary paths directly influenced the clustering properties of the fields through their sound speeds: the dilaton admitted vanishing sound speed at late times, the tachyon satisfied $c^2_{s\phi} = {-}w_\phi$, and the DBI field remained luminal ($c^2_{s\phi} \simeq 1$). 

We examined the impact of these dynamics on large-scale observables through both the (Fourier-space) linear galaxy power spectrum and the angular power spectrum. In the linear power spectrum, relativistic corrections arising from Doppler, potentials, and integrated effects, become significant only on very large scales ($k \,{\lesssim}\, 10^{-3}\, \mathrm{Mpc}^{-1}$), where they scale as powers of $\mathcal{H}/k$. As expected, these corrections enhanced the amplitude of the observed power spectrum relative to the standard (Newtonian) prediction at low $k$ and high redshift. However, we found that these relativistic contributions are largely insensitive to the choice of dark energy model: the predictions of the dilaton, the DBI field, and the cosmological constant were effectively degenerate, while the tachyon exhibited only modest deviations. This shows that the dominant relativistic terms depend mainly on geometric factors rather than on detailed dark energy physics.

Nevertheless, when comparing k-essence models directly with $\Lambda$CDM, the tachyon produced a characteristic scale-dependent signature: suppression of power on large scales and enhancement on small scales, with the amplitude of deviation increasing with redshift. Importantly, we found that neglecting relativistic corrections will induce a systematic, scale- and redshift-dependent bias in these deviations. Specifically, the omission of relativistic terms leads to an underestimation of tachyon-induced deviations at low redshifts $(z \,{\lesssim}\, 1)$ and an overestimation at higher redshifts. This highlights the need for incorporating relativistic effects when interpreting ultra-large-scale clustering data in dynamical dark energy scenarios.

Our results showed a more pronounced sensitivity to both relativistic corrections and dark energy dynamics in the angular power spectrum. Unlike the Fourier-space analysis, the full-sky treatment naturally incorporates line-of-sight integrals such as weak-lensing, ISW, and time-delay contributions, which significantly enhance the total relativistic effect. As a result, the angular power spectrum exhibited larger deviations between relativistic and standard predictions (compared to the linear power spectrum), particularly at low multipoles and high redshifts. We found that, while the dilaton and the DBI models remained degenerate with $\Lambda$CDM, the tachyon showed substantial and consistently positive deviations across all multipoles, with the magnitude increasing with redshift. This enhanced sensitivity underscores the superiority of the angular power spectrum as a probe of relativistic effects and dark energy clustering on the largest scales.

Furthermore, our analysis revealed that neglecting relativistic corrections in the angular power spectrum leads to a systematic overestimation of clustering power on large scales for all models. This contrasted with the linear power spectrum, where the bias introduced by neglecting relativistic terms varied in sign depending on the model and redshift. The robustness and consistency of the angular spectrum in capturing relativistic effects therefore make it a more reliable observable for probing deviations from $\Lambda$CDM. Also, we found that the Doppler term gives the dominant effect at most redshifts, particularly at $z < 1$, where it contributes positively and significantly to the total relativistic effect. Moreover, at intermediate redshifts ($z \sim 1$), the time-delay term becomes comparable or even dominant, while the ISW and potentials terms provide subdominant effects that grow steadily with redshift. Notably, the tachyon model consistently exhibited suppressed Doppler and time-delay contributions relative to $\Lambda$CDM, alongside enhanced ISW and potentials effects at certain epochs. These signatures provide a potential avenue for distinguishing clustering dark energy from a cosmological constant (especially when multiple relativistic contributions are jointly analysed).

Finally, the total relativistic effect in the angular power spectrum reinforces the overarching conclusion: models with near-$\Lambda$CDM background evolution (dilaton and DBI) remain observationally degenerate even when relativistic effects are included, whereas models with non-negligible deviations in $w_\phi$ (tachyon) can leave discernible imprints on ultra-large scales. However, the detectability of these signatures will depend sensitively on survey characteristics, cosmic variance, and systematic uncertainties, which were beyond the scope of this study. These results motivate the inclusion of full relativistic modeling in the analysis of forthcoming large-scale structure surveys, especially those targeting ultra-large scales and high redshifts (e.g.~Euclid~\cite{Euclid:2019clj}; LSST~\cite{LSSTDarkEnergyScience:2012kar}).

\begin{acknowledgments}
We thank the Centre for High Performance Computing, Cape Town, South Africa, for providing the computing facilities with which some of the numerical computations in this work were done. IO recieved funding from the German Academic Exchange Service (DAAD) through the In-Country/In-Region Scholarship Program. 
\end{acknowledgments}

\appendix

\section{Expanding $c^2_{s\phi}$ in $w_\phi$ (DBI)}
\label{App:DBI_wx_cs2_expand}
Here we give the Taylor expansions of both
$w_\phi(X)$ and $c^2_{s\phi}(X)$ around $X \,{=}\, 0$ (small-$X$ limit), for DBI (see \S\ref{subsec:DBI}).

\subsection{Key small-$X$ expansions}
\bea
(1+c_1 X)^{1/2} &=& 1 + \dfrac{1}{2}c_1 X - \dfrac{1}{8}c_1^2 X^2, \nn
(1+c_1 X)^{-1/2} &=& 1 - \dfrac{1}{2}c_1 X + \dfrac{3}{8}c_1^2 X^2,\nn
(1+c_1 X)^{-3/2} &=& 1 - \dfrac{3}{2}c_1 X + \dfrac{15}{8}c_1^2 X^2,\nonumber
\eea
where $(c_2X)^3 = {\cal O}(X^3)$ and $(c_3X)^4 = {\cal O}(X^4)$, so we ignore those terms, and go up to $X^2$.

\subsection{Expand $w_\phi$}
The equation of state is the ratio $w_\phi = N(X)/D(X).$ \\

\subsubsection*{\bf Numerator $N(X)$:}
\bea
N(X) &\equiv & -2.01 + 2(1+c_1 X)^{1/2},\nn
&=& -2.01 + 2\left(1 + \dfrac{1}{2}c_1 X - \dfrac{1}{8}c_1^2 X^2\right),\nn
&=& -1.01 + c_1 X - \dfrac{1}{4}c_1^2 X^2,\nonumber
\eea

\subsubsection*{\bf Denominator $D(X)$:}
\bea
D(X) &\equiv & 2.01 - 2(1+c_1 X)^{-1/2},\nn
&=& 2.01 - 2\left(1 - \dfrac{1}{2}c_1 X + \dfrac{3}{8}c_1^2 X^2\right),\nn
&=& 0.01 + c_1 X - \dfrac{3}{4}c_1^2 X^2.\nonumber
\eea
We use the expansion, $(a+bX+cX^2) / (d+eX+fX^2)$, which leads to
\beq
w_\phi = -1 + 200\,c_1 X - \left(20000 + \dfrac{1}{2}\right) c_1^2 X^2 + {\cal O}(X^3),
\eeq
where the leading behaviour is $w_\phi \,{\approx}\, {-}1 \,{+}\, 200\, c_1 X$.

\subsection{Expand $c^2_{s\phi}$:}
Note that: $c^2_{s\phi} = \partial_X N(X) / \partial_X D(X).$\\

\subsubsection*{\bf Expand $\partial_X N(X)$:}
\bea
\partial_X N(X) &=& c_1(1+c_1 X)^{-1/2},\nn
&=& c_1\left(1 - \dfrac{1}{2}c_1 X + \dfrac{3}{8}c_1^2 X^2\right),\nn
&=& c_1 - \dfrac{1}{2}c_1^2 X + \dfrac{3}{8}c_1^3 X^2.\nonumber
\eea

\subsubsection*{\bf Expand $\partial_X D(X)$:}
\bea
\partial_X D(X) &=& c_1(1+c_1 X)^{-3/2},\nn
&=& c_1\left(1 - \dfrac{3}{2}c_1 X + \dfrac{15}{8}c_1^2 X^2\right),\nn
&=& c_1 - \dfrac{3}{2}c_1^2 X + \dfrac{15}{8}c_1^3 X^2.\nonumber
\eea
From the definition of $c^2_{s\phi}$ above, after expansion:
\beq
c^2_{s\phi} = 1 + c_1 X - c_1^2 X^2 + {\cal O}(X^3),
\eeq

\subsection{Express $c^2_{s\phi}$ in terms of $w_\phi$}
We use the series we already found:
\beq\label{w_DBI}
w_\phi = -1 + 200\,c_1 X - \alpha c_1^2 X^2 + {\cal O}(X^3),
\eeq
where $\alpha = 20000 + \frac{1}{2}$, and
\beq\label{cs2_DBI}
c^2_{s\phi} = 1 + c_1 X - c_1^2 X^2 + {\cal O}(X^3).
\eeq
Let $\delta \equiv w_\phi + 1$ with $|\delta| \ll 1$, then we have
\beq\label{delta_}
\delta = 200 c_1 X - \alpha c_1^2 X^2.
\eeq

\subsubsection*{\bf Solve for $X$ (perturbatively):}
First approximation:
\beq\label{X-approx}
X \approx \dfrac{\delta}{200 c_1}.
\eeq
Next-order correction: solving iteratively, using \eqref{X-approx} in \eqref{delta_}, we have
\beq
X = \dfrac{\delta}{200 c_1} + \dfrac{\alpha c_1^2 X^2}{200 c_1} \approx \dfrac{\delta}{200 c_1} + \dfrac{\alpha}{200^3 c_1} \delta^2.
\eeq
Now we substitute into \eqref{cs2_DBI}:
\bea
c_1 X &\approx & c_1 \left(\frac{\delta}{200 c_1} + \frac{\alpha}{200^3 c_1} \delta^2\right) = \frac{\delta}{200} + \frac{\alpha}{200^3} \delta^2,\nn
c_1^2 X^2 &\approx & c_1^2 \left(\frac{\delta}{200 c_1}\right)^2 = \frac{\delta^2}{200^2} \quad (\text{keep only up to } \delta^2),\nonumber
\eea
where combining these gives
\bea
c^2_{s\phi} &=& 1 + \dfrac{\delta}{200} + \left(\dfrac{\alpha}{200^3} - \dfrac{1}{200^2}\right) \delta^2 + {\cal O}(\delta^3),\\ \nonumber
&=& 1 + \dfrac{w_\phi + 1}{200} + \hat{\alpha}\, \dfrac{(w_\phi + 1)^2}{200^2} + {\cal O}[(w_\phi + 1)^3],\nonumber
\eea
where $\hat{\alpha} \equiv \alpha/200 -1$, with $\alpha$ being as in \eqref{w_DBI}. Note that this is perturbative, only valid for $|w_\phi + 1| \,{\ll}\, 1$. For larger $w_\phi + 1$, the linear approximation can deviate significantly.


\bibliography{rels_effects_in_kessence-prd}

@PREAMBLE{
 "\providecommand{\noopsort}[1]{}" 
 # "\providecommand{\singleletter}[1]{#1}%" 
}

@article{LSSTDarkEnergyScience:2012kar,
    author = "Abate, Alexandra and others",
    collaboration = "LSST Dark Energy Science",
    title = "{Large Synoptic Survey Telescope: Dark Energy Science Collaboration}",
    eprint = "1211.0310",
    archivePrefix = "arXiv",
    primaryClass = "astro-ph.CO",
    reportNumber = "FERMILAB-FN-0952-A-T",
    doi = "10.2172/1156445",
    journal = "arXiv preprint 1211.0310",
    month = "11",
    year = "2012"
}

@article{Villa:2017yfg, 
    author = "Villa, Eleonora and Di Dio, Enea and Lepori, Francesca",
    title = "{Lensing convergence in galaxy clustering in {\ensuremath{\Lambda}}CDM and beyond}",
    eprint = "1711.07466",
    archivePrefix = "arXiv",
    primaryClass = "astro-ph.CO",
    doi = "10.1088/1475-7516/2018/04/033",
    journal = "JCAP",
    volume = "04",
    pages = "033",
    year = "2018"
}

@article{Duniya:2022miz,
    author = "Duniya, Didam G. A.",
    title = "{Qualitative probe of interacting dark energy with redshift-space distortions}",
    eprint = "2203.12582",
    archivePrefix = "arXiv",
    primaryClass = "astro-ph.CO",
    doi = "10.1142/S0218271824500123",
    journal = "Int. J. Mod. Phys. D",
    volume = "33",
    number = "02",
    pages = "2450012",
    year = "2024"
}

@article{Euclid:2019clj,
    author = "Blanchard, A. and others",
    collaboration = "Euclid",
    title = "{Euclid preparation. VII. Forecast validation for Euclid cosmological probes}",
    eprint = "1910.09273",
    archivePrefix = "arXiv",
    primaryClass = "astro-ph.CO",
    doi = "10.1051/0004-6361/202038071",
    journal = "Astron. Astrophys.",
    volume = "642",
    pages = "A191",
    year = "2020"
}

@article{Tsujikawa:2010sc,
    author = "Tsujikawa, Shinji",
    title = "{Dark energy: investigation and modeling}",
    eprint = "1004.1493",
    archivePrefix = "arXiv",
    primaryClass = "astro-ph.CO",
    doi = "10.1007/978-90-481-8685-3\_8",
    journal = "arXiv preprint 1004.1493",
    month = "4",
    year = "2010"
}

@article{Planck:2018vyg,
    author = "Aghanim, N. and others",
    collaboration = "Planck",
    title = "{Planck 2018 results. VI. Cosmological parameters}",
    eprint = "1807.06209",
    archivePrefix = "arXiv",
    primaryClass = "astro-ph.CO",
    doi = "10.1051/0004-6361/201833910",
    journal = "Astron. Astrophys.",
    volume = "641",
    pages = "A6",
    year = "2020",
    note = "[Erratum: Astron.Astrophys. 652, C4 (2021)]"
}

@article{Bamba:2012cp,
    author = "Bamba, Kazuharu and Capozziello, Salvatore and Nojiri, Shin'ichi and Odintsov, Sergei D.",
    title = "{Dark energy cosmology: the equivalent description via different theoretical models and cosmography tests}",
    eprint = "1205.3421",
    archivePrefix = "arXiv",
    primaryClass = "gr-qc",
    doi = "10.1007/s10509-012-1181-8",
    journal = "Astrophys. Space Sci.",
    volume = "342",
    pages = "155--228",
    year = "2012"
}

@Book{Amendola:2010bk,
    author = {{Amendola}, Luca and {Tsujikawa}, Shinji},
    title = {Dark Energy: Theory and Observations},
    publisher = {Cambridge University Press},
    year = {2010}
}

@article{Piazza:2004df,
    author = "Piazza, Federico and Tsujikawa, Shinji",
    title = "{Dilatonic ghost condensate as dark energy}",
    eprint = "hep-th/0405054",
    archivePrefix = "arXiv",
    reportNumber = "BICOCCA-FT-04-4",
    doi = "10.1088/1475-7516/2004/07/004",
    journal = "JCAP",
    volume = "07",
    pages = "004",
    year = "2004"
}

@article{Desjacques:2016bnm,
    author = "Desjacques, Vincent and Jeong, Donghui and Schmidt, Fabian",
    title = "{Large-Scale Galaxy Bias}",
    eprint = "1611.09787",
    archivePrefix = "arXiv",
    primaryClass = "astro-ph.CO",
    doi = "10.1016/j.physrep.2017.12.002",
    journal = "Phys. Rept.",
    volume = "733",
    pages = "1--193",
    year = "2018"
}

@article{Baldauf:2011bh,
    author = "Baldauf, Tobias and Seljak, Uros and Senatore, Leonardo and Zaldarriaga, Matias",
    title = "{Galaxy Bias and non-Linear Structure Formation in General Relativity}",
    eprint = "1106.5507",
    archivePrefix = "arXiv",
    primaryClass = "astro-ph.CO",
    doi = "10.1088/1475-7516/2011/10/031",
    journal = "JCAP",
    volume = "10",
    pages = "031",
    year = "2011"
}

@article{Bartolo:2010ec,
    author = "Bartolo, N. and Matarrese, S. and Riotto, A.",
    title = "{Relativistic Effects and Primordial Non-Gaussianity in the Galaxy bias}",
    eprint = "1011.4374",
    archivePrefix = "arXiv",
    primaryClass = "astro-ph.CO",
    reportNumber = "CERN-PH-TH-2010-270",
    doi = "10.1088/1475-7516/2011/04/011",
    journal = "JCAP",
    volume = "04",
    pages = "011",
    year = "2011"
}

@article{Bonvin:2015kuc,
    author = "Bonvin, Camille and Hui, Lam and Gaztanaga, Enrique",
    title = "{Optimising the measurement of relativistic distortions in large-scale structure}",
    eprint = "1512.03566",
    archivePrefix = "arXiv",
    primaryClass = "astro-ph.CO",
    reportNumber = "CERN-PH-TH-2015-220",
    doi = "10.1088/1475-7516/2016/08/021",
    journal = "JCAP",
    volume = "08",
    pages = "021",
    year = "2016"
}

@article{Gaztanaga:2015jrs,
    author = "Gaztanaga, Enrique and Bonvin, Camille and Hui, Lam",
    title = "{Measurement of the dipole in the cross-correlation function of galaxies}",
    eprint = "1512.03918",
    archivePrefix = "arXiv",
    primaryClass = "astro-ph.CO",
    reportNumber = "CERN-PH-TH-2015-221",
    doi = "10.1088/1475-7516/2017/01/032",
    journal = "JCAP",
    volume = "01",
    pages = "032",
    year = "2017"
}

@article{Armendariz-Picon:1999hyi,
    author = "Armendariz-Picon, C. and Damour, T. and Mukhanov, Viatcheslav F.",
    title = "{k - inflation}",
    eprint = "hep-th/9904075",
    archivePrefix = "arXiv",
    doi = "10.1016/S0370-2693(99)00603-6",
    journal = "Phys. Lett. B",
    volume = "458",
    pages = "209--218",
    year = "1999"
}

@article{Chiba:1999ka,
    author = "Chiba, Takeshi and Okabe, Takahiro and Yamaguchi, Masahide",
    title = "{Kinetically driven quintessence}",
    eprint = "astro-ph/9912463",
    archivePrefix = "arXiv",
    reportNumber = "UTAP-352",
    doi = "10.1103/PhysRevD.62.023511",
    journal = "Phys. Rev. D",
    volume = "62",
    pages = "023511",
    year = "2000"
}

@article{Armendariz-Picon:2000nqq,
    author = "Armendariz-Picon, C. and Mukhanov, Viatcheslav F. and Steinhardt, Paul J.",
    title = "{A Dynamical solution to the problem of a small cosmological constant and late time cosmic acceleration}",
    eprint = "astro-ph/0004134",
    archivePrefix = "arXiv",
    doi = "10.1103/PhysRevLett.85.4438",
    journal = "Phys. Rev. Lett.",
    volume = "85",
    pages = "4438--4441",
    year = "2000"
}

@article{Armendariz-Picon:2000ulo,
    author = "Armendariz-Picon, C. and Mukhanov, Viatcheslav F. and Steinhardt, Paul J.",
    title = "{Essentials of k essence}",
    eprint = "astro-ph/0006373",
    archivePrefix = "arXiv",
    doi = "10.1103/PhysRevD.63.103510",
    journal = "Phys. Rev. D",
    volume = "63",
    pages = "103510",
    year = "2001"
}

@article{Bamba:2011ih,
    author = "Bamba, Kazuharu and Matsumoto, Jiro and Nojiri, Shin'ichi",
    title = "{Cosmological perturbations in $k$-essence model}",
    eprint = "1109.1308",
    archivePrefix = "arXiv",
    primaryClass = "hep-th",
    doi = "10.1103/PhysRevD.85.084026",
    journal = "Phys. Rev. D",
    volume = "85",
    pages = "084026",
    year = "2012"
}

@article{Tsujikawa:2012hv,
    author = "Tsujikawa, Shinji and De Felice, Antonio and Alcaniz, Jailson",
    title = "{Testing for dynamical dark energy models with redshift-space distortions}",
    eprint = "1210.4239",
    archivePrefix = "arXiv",
    primaryClass = "astro-ph.CO",
    doi = "10.1088/1475-7516/2013/01/030",
    journal = "JCAP",
    volume = "01",
    pages = "030",
    year = "2013"
}

@article{Mulki:2020hwt,
    author = "Mulki, Fargiza A. M. and Wulandari, Hesti",
    editor = "Ng, W. K. and Yang, A. and Ng, S. C. C. and Chan, A. H. and Sow, C. H.",
    title = "{Analytical treatment of small scales matter power spectrum in coupled scalar field (CSF) cosmology}",
    doi = "10.1051/epjconf/202024002003",
    journal = "EPJ Web Conf.",
    volume = "240",
    pages = "02003",
    year = "2020"
}

@article{Mohammadi:2019qeu,
    author = "Mohammadi, Abolhassan and Golanbari, Tayeb and Saaidi, Khaled",
    title = "{Beta-function formalism for k-essence constant-roll inflation}",
    eprint = "1912.07006",
    archivePrefix = "arXiv",
    primaryClass = "gr-qc",
    doi = "10.1016/j.dark.2020.100505",
    journal = "Phys. Dark Univ.",
    volume = "28",
    pages = "100505",
    year = "2020"
}

@article{Rezazadeh:2020zrd,
    author = "Rezazadeh, K. and Asadzadeh, S. and Fahimi, K. and Karami, K. and Mehrabi, A.",
    title = "{The growth of DM and DE perturbations in DBI non-canonical scalar field scenario}",
    eprint = "2001.07920",
    archivePrefix = "arXiv",
    primaryClass = "gr-qc",
    doi = "10.1016/j.aop.2020.168299",
    journal = "Annals Phys.",
    volume = "422",
    pages = "168299",
    year = "2020"
}

@article{Brando:2020ouk,
    author = "Brando, Guilherme and Koyama, Kazuya and Wands, David",
    title = "{Relativistic Corrections to the Growth of Structure in Modified Gravity}",
    eprint = "2006.11019",
    archivePrefix = "arXiv",
    primaryClass = "astro-ph.CO",
    doi = "10.1088/1475-7516/2021/01/013",
    journal = "JCAP",
    volume = "01",
    pages = "013",
    year = "2021"
}

@article{Duniya:2022vdi,
    author = "Duniya, Didam G. A. and Abebe, Amare and de la Cruz-Dombriz, Alvaro and Dunsby, Peter K. S.",
    title = "{Imprint of f(R) gravity in the cosmic magnification}",
    eprint = "2210.09303",
    archivePrefix = "arXiv",
    primaryClass = "astro-ph.CO",
    doi = "10.1093/mnras/stac3538",
    journal = "Mon. Not. Roy. Astron. Soc.",
    volume = "518",
    number = "4",
    pages = "6102--6113",
    year = "2022"
}

@article{Duniya:2023xgx,
    author = "Duniya, Didam and Mongwane, Bishop",
    title = "{Cosmic magnification in beyond-Horndeski gravity}",
    eprint = "2311.04169",
    archivePrefix = "arXiv",
    primaryClass = "astro-ph.CO",
    journal = "\href{https://arxiv.org/abs/2311.04169}{arXiv:2311.04169}",
    month = "11",
    year = "2023"
}

@article{Duniya:2016gcf,
    author = "Duniya, Didam",
    title = "{Large-scale imprint of relativistic effects in the cosmic magnification}",
    eprint = "1604.03934",
    archivePrefix = "arXiv",
    primaryClass = "astro-ph.CO",
    doi = "10.1103/PhysRevD.93.103538",
    journal = "Phys. Rev. D",
    volume = "93",
    number = "10",
    pages = "103538",
    year = "2016",
    note = "[Addendum: Phys.Rev.D 93, 129902 (2016)]"
}

@article{Yoo:2009au,
    author = "Yoo, Jaiyul and Fitzpatrick, A. Liam and Zaldarriaga, Matias",
    title = "{A New Perspective on Galaxy Clustering as a Cosmological Probe: General Relativistic Effects}",
    eprint = "0907.0707",
    archivePrefix = "arXiv",
    primaryClass = "astro-ph.CO",
    doi = "10.1103/PhysRevD.80.083514",
    journal = "Phys. Rev. D",
    volume = "80",
    pages = "083514",
    year = "2009"
}

@article{Yoo:2010ni,
    author = "Yoo, Jaiyul",
    title = "{General Relativistic Description of the Observed Galaxy Power Spectrum: Do We Understand What We Measure?}",
    eprint = "1009.3021",
    archivePrefix = "arXiv",
    primaryClass = "astro-ph.CO",
    doi = "10.1103/PhysRevD.82.083508",
    journal = "Phys. Rev. D",
    volume = "82",
    pages = "083508",
    year = "2010"
}

@article{Yoo:2014kpa,
    author = "Yoo, Jaiyul",
    title = "{Relativistic Effect in Galaxy Clustering}",
    eprint = "1409.3223",
    archivePrefix = "arXiv",
    primaryClass = "astro-ph.CO",
    doi = "10.1088/0264-9381/31/23/234001",
    journal = "Class. Quant. Grav.",
    volume = "31",
    pages = "234001",
    year = "2014"
}

@article{Challinor:2011bk,
    author = "Challinor, Anthony and Lewis, Antony",
    title = "{The linear power spectrum of observed source number counts}",
    eprint = "1105.5292",
    archivePrefix = "arXiv",
    primaryClass = "astro-ph.CO",
    doi = "10.1103/PhysRevD.84.043516",
    journal = "Phys. Rev. D",
    volume = "84",
    pages = "043516",
    year = "2011"
}

@article{Lopez-Honorez:2011emg,
    author = "Lopez-Honorez, L. and Mena, O. and Rigolin, S.",
    title = "{Biases on cosmological parameters by general relativity effects}",
    eprint = "1109.5117",
    archivePrefix = "arXiv",
    primaryClass = "astro-ph.CO",
    reportNumber = "DFPD-11-TH-13, IFIC-11-51, ULB-TH-11-24",
    doi = "10.1103/PhysRevD.85.023511",
    journal = "Phys. Rev. D",
    volume = "85",
    pages = "023511",
    year = "2012"
}

@article{Alonso:2015uua,
    author = "Alonso, David and Bull, Philip and Ferreira, Pedro G. and Maartens, Roy and Santos, M.",
    title = "{Ultra large-scale cosmology in next-generation experiments with single tracers}",
    eprint = "1505.07596",
    archivePrefix = "arXiv",
    primaryClass = "astro-ph.CO",
    doi = "10.1088/0004-637X/814/2/145",
    journal = "Astrophys. J.",
    volume = "814",
    number = "2",
    pages = "145",
    year = "2015"
}

@article{Bonvin:2014owa,
    author = "Bonvin, Camille",
    title = "{Isolating relativistic effects in large-scale structure}",
    eprint = "1409.2224",
    archivePrefix = "arXiv",
    primaryClass = "astro-ph.CO",
    doi = "10.1088/0264-9381/31/23/234002",
    journal = "Class. Quant. Grav.",
    volume = "31",
    number = "23",
    pages = "234002",
    year = "2014"
}

@article{Durrer:2016jzq,
    author = "Durrer, Ruth and Tansella, Vittorio",
    title = "{Vector perturbations of galaxy number counts}",
    eprint = "1605.05974",
    archivePrefix = "arXiv",
    primaryClass = "astro-ph.CO",
    doi = "10.1088/1475-7516/2016/07/037",
    journal = "JCAP",
    volume = "07",
    pages = "037",
    year = "2016"
}

@article{Duniya:2022xcz,
    author = "Duniya, Didam G. A. and Kumwenda, Mazuba",
    title = "{Which is a better cosmological probe: number counts or cosmic magnification?}",
    eprint = "2203.11159",
    archivePrefix = "arXiv",
    primaryClass = "astro-ph.CO",
    doi = "10.1093/mnras/stad1231",
    journal = "Mon. Not. Roy. Astron. Soc.",
    volume = "522",
    number = "3",
    pages = "3308--3317",
    year = "2023"
}

@article{Clifton:2011jh,
    author = "Clifton, Timothy and Ferreira, Pedro G. and Padilla, Antonio and Skordis, Constantinos",
    title = "{Modified Gravity and Cosmology}",
    eprint = "1106.2476",
    archivePrefix = "arXiv",
    primaryClass = "astro-ph.CO",
    doi = "10.1016/j.physrep.2012.01.001",
    journal = "Phys. Rept.",
    volume = "513",
    pages = "1--189",
    year = "2012"
}

@article{Ishak:2018his,
    author = "Ishak, Mustapha",
    title = "{Testing General Relativity in Cosmology}",
    eprint = "1806.10122",
    archivePrefix = "arXiv",
    primaryClass = "astro-ph.CO",
    doi = "10.1007/s41114-018-0017-4",
    journal = "Living Rev. Rel.",
    volume = "22",
    number = "1",
    pages = "1",
    year = "2019"
}

@article{Duniya:2019mpr,
    author = "Duniya, Didam and Moloi, Teboho and Clarkson, Chris and Larena, Julien and Maartens, Roy and Mongwane, Bishop and Weltman, Amanda",
    title = "{Probing beyond-Horndeski gravity on ultra-large scales}",
    eprint = "1902.09919",
    archivePrefix = "arXiv",
    primaryClass = "astro-ph.CO",
    doi = "10.1088/1475-7516/2020/01/033",
    journal = "JCAP",
    volume = "01",
    pages = "033",
    year = "2020"
}

@article{Duniya:2013eta,
    author = "Duniya, Didam and Bertacca, Daniele and Maartens, Roy",
    title = "{Clustering of quintessence on horizon scales and its imprint on HI intensity mapping}",
    eprint = "1305.4509",
    archivePrefix = "arXiv",
    primaryClass = "astro-ph.CO",
    doi = "10.1088/1475-7516/2013/10/015",
    journal = "JCAP",
    volume = "10",
    pages = "015",
    year = "2013"
}

@article{Duniya:2015nva,
    author = "Duniya, Didam G. A. and Bertacca, Daniele and Maartens, Roy",
    title = "{Probing the imprint of interacting dark energy on very large scales}",
    eprint = "1502.06424",
    archivePrefix = "arXiv",
    primaryClass = "astro-ph.CO",
    doi = "10.1103/PhysRevD.91.063530",
    journal = "Phys. Rev. D",
    volume = "91",
    pages = "063530",
    year = "2015"
}

@article{Duniya:2015dpa,
    author = "Duniya, Didam",
    title = "{Dark energy homogeneity in general relativity: Are we applying it correctly?}",
    eprint = "1505.03436",
    archivePrefix = "arXiv",
    primaryClass = "gr-qc",
    doi = "10.1007/s10714-016-2047-0",
    journal = "Gen. Rel. Grav.",
    volume = "48",
    number = "4",
    pages = "52",
    year = "2016"
}

@article{Duniya:2016ibg,
    author = "Duniya, Didam",
    title = "{Understanding the relativistic overdensity of galaxy surveys}",
    eprint = "1606.00712",
    archivePrefix = "arXiv",
    primaryClass = "astro-ph.CO",
    journal = "\href{https://arxiv.org/abs/1606.00712}{arXiv:1606.00712}",
    month = "5",
    year = "2016"
}

@article{Jeong:2011as,
    author = "Jeong, Donghui and Schmidt, Fabian and Hirata, Christopher M.",
    title = "{Large-scale clustering of galaxies in general relativity}",
    eprint = "1107.5427",
    archivePrefix = "arXiv",
    primaryClass = "astro-ph.CO",
    doi = "10.1103/PhysRevD.85.023504",
    journal = "Phys. Rev. D",
    volume = "85",
    pages = "023504",
    year = "2012"
}

@article{Bardeen:1980kt,
    author = "Bardeen, James M.",
    title = "{Gauge Invariant Cosmological Perturbations}",
    doi = "10.1103/PhysRevD.22.1882",
    journal = "Phys. Rev. D",
    volume = "22",
    pages = "1882--1905",
    year = "1980"
}

@article{Bonvin:2011bg,
    author = "Bonvin, Camille and Durrer, Ruth",
    title = "{What galaxy surveys really measure}",
    eprint = "1105.5280",
    archivePrefix = "arXiv",
    primaryClass = "astro-ph.CO",
    doi = "10.1103/PhysRevD.84.063505",
    journal = "Phys. Rev. D",
    volume = "84",
    pages = "063505",
    year = "2011"
}

\end{document}